\documentclass[pre,twocolumn,nobalancelastpage,amsmath,amssymb,showpacs,
nofootinbib]{revtex4-1}

\usepackage{latexsym}
\usepackage{graphics}
\usepackage{xcolor}
\usepackage{mathtools}
\usepackage{amsmath,amssymb}

\begin{document}
\title{Athermal shearing of frictionless cross-shaped particles of varying aspect ratio}
\author{Theodore A. Marschall and S. Teitel}

\affiliation{Department of Physics and Astronomy, University of Rochester, Rochester, NY 14627}
\date{today}

\begin{abstract}
We use numerical simulations to study the shear-driven steady-state flow of athermal, frictionless, overdamped, two dimensional cross-shaped particles of varying aspect ratios, and make comparison with the behavior of rod-shaped and staple-shaped particles.  We find that the extent of non-convexity of the particle shape plays an important role in determining both the value of the jamming packing fraction as well as the rotational motion and orientational ordering of the particles.

\end{abstract}
\maketitle
\section{Introduction}
\label{intro}

Models of athermal ($T=0$) particles, interacting through soft-core repulsive interactions, have been used to study a wide variety of granular systems, such as dry granular materials, foams, emulsions, and non-Brownian suspensions. Most such works have focused on the simplest case of spherical particles.  More recently, attention has been paid to the case of aspherical particles with lower rotational symmetry, such as rods or  ellipsoids \cite{Borzsonyi.Soft.2013}.  Relatively few works have considered particles with a non-convex shape  \cite{Reddy1,Reddy2,Saint-Cyr,Guo1,Jaeger,Zheng,Mandal,VanderWerf,MarschallStaples}.  
{\color{black}A particle is non-convex if there exists two points on the surface such that the cord  connecting the two points does not lie entirely inside the particle.}
In this work we consider the flow of cross-shaped particles in two dimensions (2D) driven by steady-state simple shear at a constant strain rate $\dot\gamma$.  The non-convex shape of the crosses allows for particles to interlock and create gear-like effects in their interactions.   We study the rotational motion of such particles and their orientational ordering in the shear flow, making comparison to previous work we have done  on non-convex U-shaped particles (``staples") \cite{MarschallStaples} and convex elongated rods \cite{MKOT,MT1,MT2}.  We will see that the lack of convexity plays a significant role in such particle orientational effects, 
{\color{black}affecting whether the average angular velocity and the degree of orientation ordering increases or decreases, as the packing varies.  We will also find a surprising linear relation between the packing fraction at jamming $\phi_J$ and the degree of non-convexity of the particle shape.}

\section{Model}
\label{model}

\begin{figure}
\resizebox{0.95\hsize}{!}{
\includegraphics{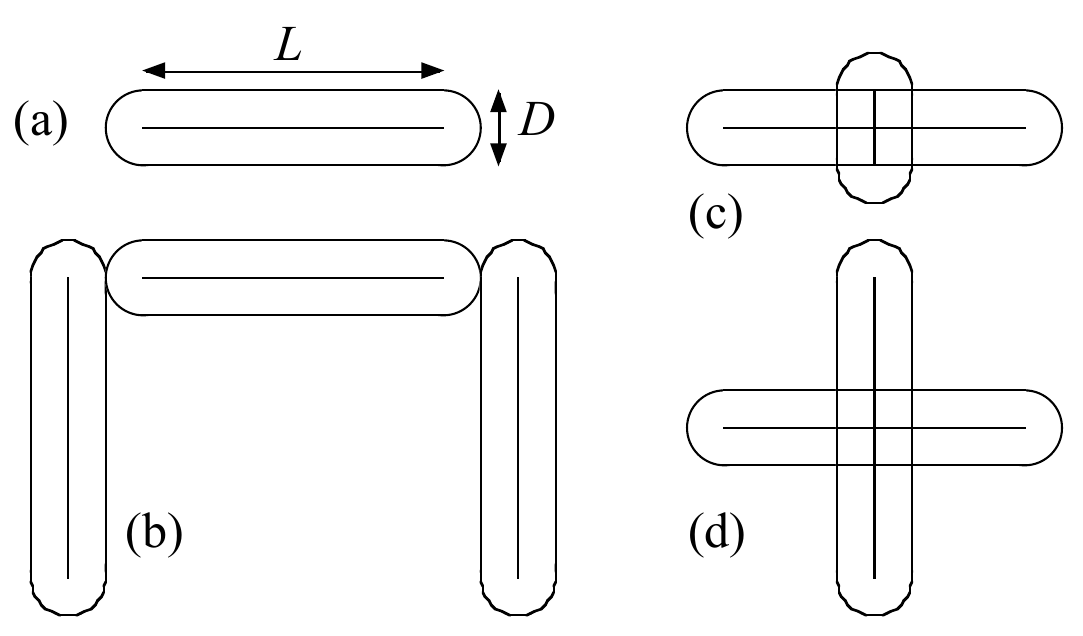}}
\caption{(a) Spherocylinder in two dimensions with spine of length $L$ and width $D$ and asphericity $\alpha=L/D=4$; (b) Staple formed by rigidly attaching three spherocylinders together; (c) and (d) Crosses formed by overlapping spherocylinders: (c) has aspect ratio $\beta=L^\prime/L=0.25$ while (d) has aspect ratio $\beta=1$.}
\label{shapes}
\end{figure}

\begin{figure}
\resizebox{0.95\hsize}{!}{
\includegraphics{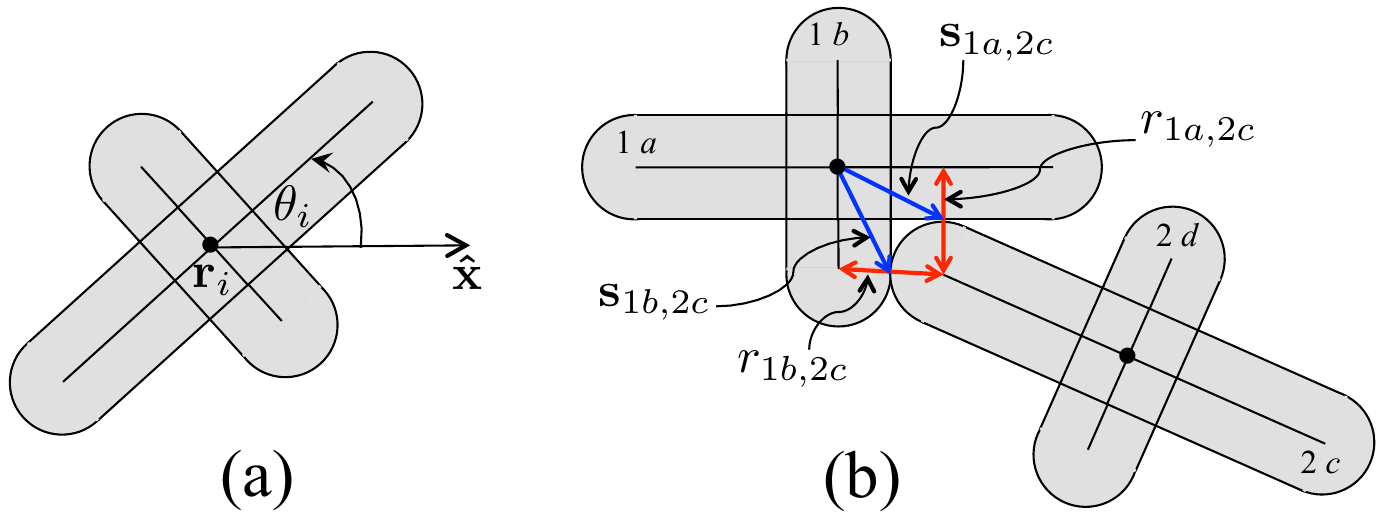}}
\caption{(a) Sketch of the degrees of  freedom of a cross: the center of mass position $\mathbf{r}_i$ and the angle $\theta_i$ of the long arm spine with respect to the shear flow direction $\mathbf{\hat x}$.  The angle $\theta_i$ increases as the cross rotates counter-clockwise.
(b) Sketch of two crosses with a pair of contacts; spherocylinder arm $c$ of cross 2 makes contact with both arms $a$ and $b$ of cross 1.  The elastic forces at these two contacts are determined by the lengths $r_{1a,2c}$ and $r_{1b,2c}$.  Also shown are the moment arms $\mathbf{s}_{1a,2c}$ and $\mathbf{s}_{1b,2c}$ of the torques that act on cross 1 from the two contacts. 
}
\label{cross-contact}
\end{figure}

The basic building block of our particles is a spherocylinder, which in 2D consists of a rectangle of length $L$ and width $D$, capped by semi-circular end caps of diameter $D$, as shown in Fig.~\ref{shapes}a.  We define the asyphericity of the spherocylinder as $\alpha=L/D$, such that $\alpha=0$ is a pure circle.  We refer to the line that bisects the rectangle parallel to its length as the ``spine" of the spherocylinder.  The shortest distance from the spine to any point on the surface is always $D/2$.  Staples are made by rigidly fixing three equal spherocylinders together as shown in Fig.~\ref{shapes}b.  Crosses are made by fixing together two orthogonal spherocylinders with overlapping centers of mass, as in Figs.~\ref{shapes}c,d.  For an isolated spherocylinder, as well as the three component spherocylinders of a staple, and the long arm of a cross, we use spherocylinders with fixed $\alpha=4$.  For the short arm of the cross we use spherocylinders with different values of $L^\prime/D$, where the width $D$ is the same as for the long arm.  We define the  aspect ratio of the cross as $\beta=L^\prime/L$.  
{\color{black}We will specify the degrees of freedom of a cross by its center of mass position $\mathbf{r}_i$, and the angular orientation $\theta_i$ of the spine of the long arm with respect to the shear flow direction $\mathbf{\hat x}$, as indicated in Fig.~\ref{cross-contact}a.}

%For the isolated spherocylinders and the crosses we take a bidisperse distribution of particle sizes, with equal numbers of big and small particles with length scales in the ratio $L_b/L_s=1.4$.  The staples are monodisperse in size.

If $ia$ labels spherocylinder component $a$ of particle $i$, and $jb$ labels spherocylinder component $b$ of particle $j$, 
we define $r_{ia,jb}$ as the shortest distance between the spines of $ia$ and $jb$. Particles $i$ and $j$ will overlap if $r_{ia,jb} <D_{ij}= (D_i+D_j)/2$.  In that case there is a harmonic repulsion between the particles with an elastic force on particle $i$ \cite{MarschallStaples,MT1},
\begin{equation}
\mathbf{F}_{ia,jb}^\mathrm{el}=\frac{k_e}{D_{ij}}\left(1-\frac{r_{ia,jb}}{D_{ij}}\right)\mathbf{\hat n}_{ia,jb}
\end{equation}
where $k_e$ is the soft-core stiffness and $\mathbf{\hat n}_{ia,jb}$ is a unit normal pointing inward to spherocylinder $ia$ at the point of contact with $jb$.  For non-convex particles, such as the crosses, a given pair of particles $i$ and $j$ may have more than one mutual contact, 
{\color{black}as illustrated in Fig.~\ref{cross-contact}b}.  
The total elastic force on particle $i$ is,
\begin{equation}
\mathbf{F}_i^\mathrm{el}=\sum_{a,jb}\mathbf{F}_{ia,jb}^\mathrm{el}
\end{equation}
where the sum is over all contacts that $i$ makes with other particles $j$.   These elastic forces also exert a torque on particle $i$.  The torque about the center of mass $\mathbf{r}_i$ of particle $i$ is, 
\begin{equation}
\tau_i^\mathrm{el}=\mathbf{\hat z}\cdot\sum_{a,jb} \mathbf{s}_{ia,jb}\times \mathbf{F}_{ia,jb}^\mathrm{el}
\end{equation}
where $\mathbf{s}_{ia,jb}$ is the moment arm from the center of mass $\mathbf{r}_i$ of particle $i$ to the point of contact between spherocylinder component $a$ of particle $i$ with spherocylinder component $b$ of particle $j$; $\mathbf{\hat z}$ is normal to the plane of the particles.

Shearing  inputs energy into the system and so there must be a mechanism for energy dissipation if a steady-state is to be reached.  Here we will assume that this dissipation occurs via a drag force with respect to a uniformly sheared host medium, thus modeling an emulsion or non-Brownian suspension \cite{MarschallStaples,MT1}.  Taking the local velocity of this host medium as  a uniform shear flow in the $\mathbf{\hat x}$ direction,
\begin{equation}
\mathbf{v}_\mathrm{host}(\mathbf{r})=\dot\gamma y\mathbf{\hat x},
\end{equation}
then gives a drag force density 
\begin{equation}
\mathbf{f}_i^\mathrm{dis}(\mathbf{r})=-k_d[\mathbf{v}_i(\mathbf{r})-\mathbf{v}_\mathrm{host}(\mathbf{r})]
\end{equation}
where
\begin{equation}
\mathbf{v}_i(\mathbf{r})=\mathbf{\dot r}_i+\dot\theta_i\mathbf{\hat z}\times(\mathbf{r}-\mathbf{r}_i)
\end{equation}
is the local velocity at position $\mathbf{r}$ on the particle $i$.  Here $\mathbf{\dot r}_i$ is the center of mass velocity of the particle, and $\dot\theta_i\mathbf{\hat z}$ is the angular velocity about the center of mass.  We measure the orientation of a particle by the angle $\theta_i$ that is made between the flow direction $\mathbf{\hat x}$ and (i) the spine of an isolated spherocylinder as in Fig.~\ref{shapes}a, (ii) the cross piece connecting the two prongs of the staple as in Fig.~\ref{shapes}b, and (iii) the long arm of the cross as in Figs.~\ref{shapes}c,d.

The total dissipative force on particle $i$ is then,
\begin{equation}
\mathbf{F}_i^\mathrm{dis}=\int_i d^2r\, \mathbf{f}_i^\mathrm{dis}(\mathbf{r}),
\end{equation}
while the total dissipative torque is,
\begin{equation}
\tau_i^\mathrm{dis}=\mathbf{\hat z}\cdot\int_i d^2r\, (\mathbf{r}-\mathbf{r}_i)\times \mathbf{f}_i^\mathrm{dis}(\mathbf{r}),
\end{equation}
where the integrals are over the area of particle $i$.

Assuming a uniform mass density for the particle, then $\int_i d^2r\,(\mathbf{r}-\mathbf{r}_i)=0$ by the definition of the center of  mass, and since $\mathbf{v}_\mathrm{host}(\mathbf{r})$ is linear in $\mathbf{r}$, one can show that the dissipative force on $i$ reduces to,
\begin{equation}
\mathbf{F}_i^\mathrm{dis}=-k_d\mathcal{A}_i[\mathbf{\dot r}_i-\dot\gamma y\mathbf{\hat x}],
\end{equation}
where $\mathcal{A}_i$ is the area of particle $i$.  The dissipative torque on $i$ can be shown to reduce to \cite{MarschallStaples,MT1},
\begin{equation}
\tau_i^\mathrm{dis}=-k_d\mathcal{A}_i I_i[\dot\theta_i+\dot\gamma f(\theta_i)],
\end{equation}
where 
\begin{equation}
f(\theta)=\frac{1}{2}[1-(\Delta I_i/I_i)\cos 2\theta].
\label{eftheta}
\end{equation}
Here  $I_i$ is the sum of the two eigenvalues of the moment of inertia tensor of particle $i$ while $\Delta I_i$ is their difference.  

We will use an overdamped dynamics (i.e., limit of small particle mass) for which,
\begin{equation}
\mathbf{F}_i^\mathrm{el}+\mathbf{F}_i^\mathrm{dis}=0,\qquad
\tau_i^\mathrm{el}+\tau_i^\mathrm{dis}=0.
\end{equation}
This leads to equations for the translational and rotation motion of particle $i$ \cite{MarschallStaples,MT1},
%\begin{equation}
%\begin{array}{ll}
%\begin{align}
\begin{eqnarray}
\mathbf{\dot r}_i &= \dot\gamma y_i\mathbf{\hat x}+\displaystyle{\frac{\mathbf{F}_i^\mathrm{el}}{k_d\mathcal{A}_i}},
\label{erdot}\\[10pt]
\dot\theta_i&=-\dot\gamma f(\theta_i)+\displaystyle{\frac{\tau_i^\mathrm{el}}{k_dI_i\mathcal{A}_i}}.
\label{ethetadot}
\end{eqnarray}
%\end{align}
%\end{array}
%\end{equation}
The packing fraction $\phi$ of our system of $N$ particles is,
\begin{equation}
\phi = \frac{1}{\mathcal{L}^2}\sum_{i=1}^N\mathcal{A}_i,
\end{equation}
where $\mathcal{L}$ is the length of our system box in both $\mathbf{\hat x}$ and $\mathbf{\hat y}$ directions.

We simulate an ensemble of $N=512$ crosses, monodisperse in both size and aspect ratio $\beta$, considering the particular cases $\beta=0.25$, 0.5 and 1.  
We  take $D = 1$ as the unit of length, $k_e = 1$ as the unit of energy, and $t_0 = D^2 k_d / k_e = 1$ as the unit of time.
We numerically integrate the equations of motion (\ref{erdot}) and (\ref{ethetadot}) using a two-stage Heun method with a step size of $\Delta t = 0.02$.
We implement a uniform simple shear using periodic boundary conditions in the $\mathbf{\hat x}$ direction, and Lees-Edwards boundary conditions \cite{LeesEdwards} in the $\mathbf{\hat y}$ direction, with shear strain $\gamma=\dot\gamma t$ for constant $\dot\gamma t_0=10^{-4}$, $10^{-5}$, and $10^{-6}$.  Simulations are started from independent initial configurations at each value of $\phi$ and $\dot\gamma$, with particles placed so as to exclude the crossing of spines belonging to different particles, but otherwise  with random positions and orientations.  We shear to a total strain $\gamma_\mathrm{max}\sim 100 - 120$, with an initial strain of $\gamma\sim10$ discarded from our ensemble averages so as to reach the steady-state.

\section{Observables}
\label{observables}

We are interested in the rotational motion of the particles, which is driven by the $\dot\gamma f(\theta_i)$ term in Eq.~(\ref{ethetadot}).  We thus measure the average angular velocity, scaled by the strain rate,
\begin{equation}
\langle\dot\theta_i\rangle/\dot\gamma = \left\langle \frac{1}{N\dot\gamma}\sum_{i=1}^N \dot\theta_i\right\rangle,
\end{equation}
where $\langle\dots\rangle$ indicates an average over configurations in the steady-state.
 
We are also interested in the orientational ordering of the particles.  For a 2D system the magnitude $S_m$ and direction $\theta_m$ of the $m$-fold orientational order parameter $\mathbf{S}_m$ are given by \cite{Donev},
\begin{equation}
S_m=\max_{\theta_m}\left[\left\langle \frac{1}{N}\sum_{i=1}^N\cos (m[\theta_i-\theta_m])\right\rangle\right],
\end{equation}
from which one can show \cite{Donev},
\begin{equation}
S_m=\sqrt{\left\langle\frac{1}{N}\sum_{i=1}^N \cos (m\theta_i)\right\rangle^2
+\left\langle\frac{1}{N}\sum_{i=1}^N \sin (m\theta_i)\right\rangle^2}
\end{equation}
and
%\begin{equation}
%\tan (m\theta_m) = \displaystyle{\frac
%{\left\langle \displaystyle{\frac{1}{N}\sum_{i=1}^N\sin (m\theta_i)}\right\rangle}
%{\left\langle \displaystyle{\frac{1}{N}\sum_{i=1}^N\cos (m\theta_i)}\right\rangle}
%}.
%\end{equation}
\begin{equation}
\tan (m\theta_m) = 
{\left\langle \displaystyle{\frac{1}{N}\sum_{i=1}^N\sin (m\theta_i)}\right\rangle}\bigg/
{\left\langle \displaystyle{\frac{1}{N}\sum_{i=1}^N\cos (m\theta_i)}\right\rangle}
.
\end{equation}

\section{Isolated Particles}
\label{isolated}

Note, for an {\em isolated} particle, where $\tau_i^\mathrm{el}=0$, rotational motion is given simply by the deterministic equation $\dot\theta_i=-\dot\gamma f(\theta_i)$, with $f(\theta)$ as in Eq.~(\ref{eftheta}).  The particle will rotate continuously clockwise, but with a non-uniform angular velocity that is slowest at $\theta_i=0$ or $\pi$ where $f(\theta_i)$ is at its minimum, and fastest at $\theta_i=\pi/2$ or  $3\pi/2$ where $f(\theta_i)$ is at its maximum.  The particle will thus spend more time oriented at $\theta_i=0$, aligned parallel to the flow direction $\mathbf{\hat x}$.

In this isolated particle limit one finds that the probability for the particle to be at angle $\theta_i\in[0,2\pi)$ is \cite{MarschallStaples},
\begin{equation}
\mathcal{P}(\theta_i)=\displaystyle{\frac{\sqrt{1-C^2}}
%{4\pi f(\theta)}}
{2\pi[1-C\cos( 2\theta_i)]}},
\end{equation}
where $C=\Delta I_i/I_i$ \cite{MT1}.
This gives for the average angular velocity \cite{MarschallStaples},
\begin{equation}
-\langle\dot\theta_i\rangle/\dot\gamma = \int_0^{2\pi}\!\!\! d\theta\,\mathcal{P}(\theta) f(\theta)=
\frac{1}{2}\sqrt{1-C^2}  < 1/2.
\label{eomega}
\end{equation}
Since $\mathcal{P}(\theta)$ peaks at $\theta=0$, one has $\theta_2=0$ for the orientation of the nematic director, while the magnitude of the nematic order parameter is given by,
\begin{equation}
S_2=\int_0^{2\pi}\!\!\! d\theta\,\mathcal{P}(\theta)\cos(2\theta)
=\displaystyle{
\frac{1-\sqrt{1-C^2}}{C}
}
\label{eS2}
\end{equation}

For crosses for which the long arm is a spherocylinder of asphericity $\alpha \ge 1$, and for which the  aspect ratio $\beta$ satisfies $1/\alpha \le \beta\le 1$, we have,
\begin{equation}
\scalebox{.93}[1]
{$
%\displaystyle{\frac{\Delta I}{I}
%}
C
=
\displaystyle{\frac{4\alpha(1-\beta)+3\pi\alpha^2(1-\beta^2)+4\alpha^3(1-\beta^3)}
{3\pi - 8 +12\alpha(1+\beta)+3\pi\alpha^2(1+\beta^2)+4\alpha^3(1+\beta^3)}
}.
$}
\label{ebeta}
\end{equation}

\begin{figure}
\resizebox{0.95\hsize}{!}{
\includegraphics{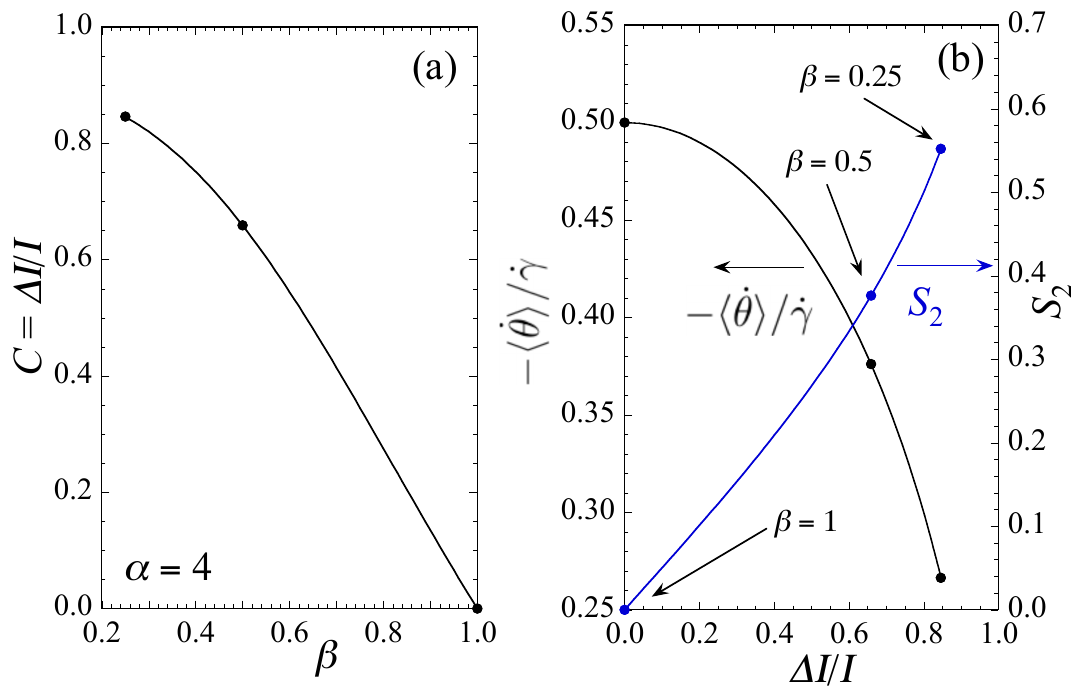}}
\caption{(a) Relation between $C=\Delta I/I$ and cross aspect ratio $\beta$, for crosses in which the asphericity of the long arm is $\alpha=4$, from Eq.~(\ref{ebeta}). (b) Scaled average particle angular velocity $-\langle\dot\theta\rangle/\dot\gamma$ and magnitude of the nematic order parameter $S_2$ vs $\Delta I/I$, for an isolated particle, from Eqs.~(\ref{eomega}) and (\ref{eS2}).  In both panels the solid points indicate the three values of $\beta=0.25$, 0.5 and 1 where we have done numerical simulations.}
\label{single}
\end{figure}

In Fig.~\ref{single}a we plot $C=\Delta I/I$ vs $\beta$ for crosses in which the asphericity of the long arm is $\alpha=4$.  In Fig.~\ref{single}b we plot $-\langle \dot\theta\rangle/\dot\gamma$ and $S_2$ vs $\Delta I/I$ for an isolated particle.  We see that as $\Delta I/I\to 0$, $-\langle\dot\theta\rangle/\dot\gamma\to 1/2$ and $S_2\to 0$, as expected for a circular particle.

At finite packing $\phi$, particles will come into contact, $\tau_i^\mathrm{el}$ will no longer be zero, and the above isolated particle behavior will be modified.  The goal of this work is to see how the non-convex shape of the crosses influences the effect of this $\tau_i^\mathrm{el}$.

A particularly interesting case is the cross with aspect ratio $\beta=1$, i.e., arms of equal length.  In this case the $4$-fold rotational symmetry of the particle results in $\Delta I_i=0$, and so such an isolated cross rotates uniformly with $-\dot\theta/\dot\gamma = 1/2$, and the probability distribution $\mathcal{P}(\theta_i)=1/2\pi$ is completely uniform, just like for a circular particle.  There is thus no orientational ordering for an isolated cross with $\beta=1$, and in this limit $S_m=0$ for all $m$.  At finite packing $\phi$, any orientational ordering that is observed is necessarily due to the particle collisions and the resulting $\tau_i^\mathrm{el}$.  In such a case, the rotational symmetry of the cross still necessarily results in a nematic order parameter $S_2=0$, and one must look to the tetratic $S_4$ for indications of orientational ordering.

\section{Numerical Results}
\label{results}

\subsection{The Jamming Transition}
\label{jamming}

Before considering the rotational and orientational behavior of a finite density of  crosses, we first digress to look at the location of the jamming transition $\phi_J$.  To look for the jamming transition we measure the pressure $p$ 
{\color{black}due to the particle interactions (we ignore any ambient pressure of the host medium)} 
which is 1/2 the trace of the stress tensor \cite{MarschallStaples},
\begin{equation}
\mathbf{P}=-\frac{1}{\mathcal{L}^2}\sum_i \sum_{a,jb}\mathbf{s}_{ia,jb}\otimes \mathbf{F}_{ia,jb}^\mathrm{el}.
\label{eP}
\end{equation}
Here the second sum is over all contacts between spherocylinder component $a$ of particle $i$ with spherocylinder $b$ of particle $j$, and we consider only the elastic forces since these give the dominant contribution to the stress at low $\dot\gamma$.  

\begin{figure}
\resizebox{0.95\hsize}{!}{
\includegraphics{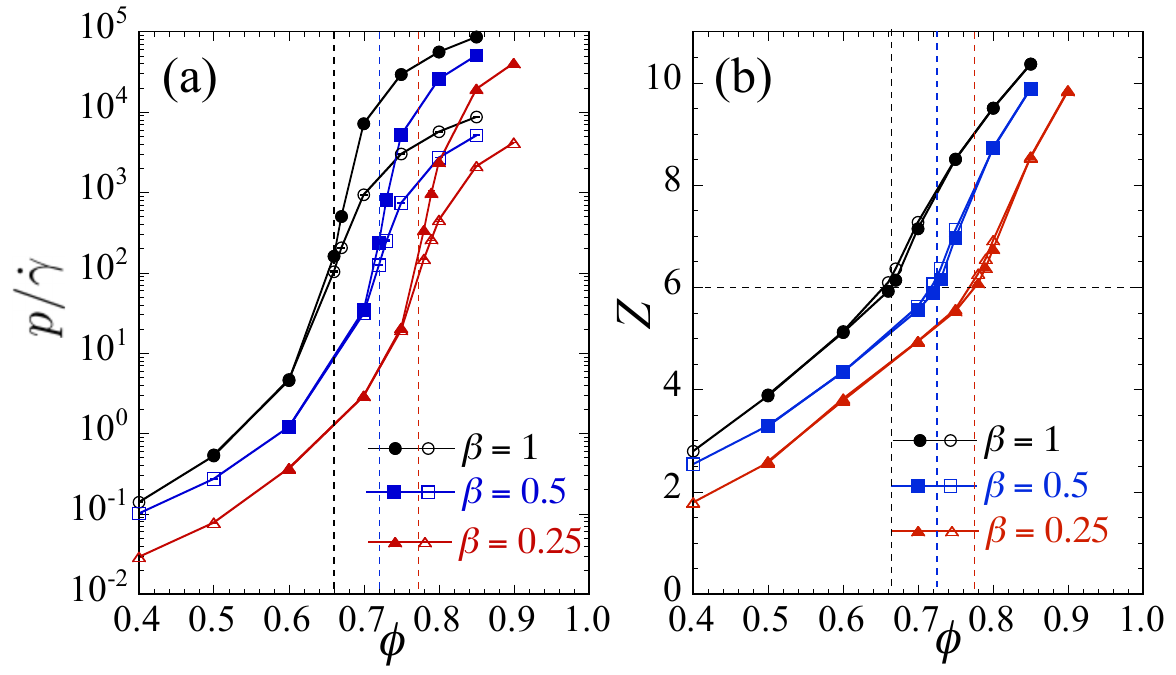}}
\caption{For crosses with aspect ratios $\beta=0.25$, 0.5 and 1,  (a) ratio of pressure to strain rate $p/\dot\gamma$ vs $\phi$, and (b) average number of contacts per particle $Z$ vs $\phi$.  In both panels, solid symbols denote data at $\dot\gamma=10^{-6}$, while open symbols denote data at $\dot\gamma=10^{-5}$. Dashed vertical lines in (a) denote the lower bound estimate of $\phi_J$, while in (b) they denote the upper bound estimate of $\phi_J$ as given by the packing at which $Z=Z_\mathrm{iso}=6$.
}
\label{eta-vs-phi}
\end{figure}

For our overdamped model with a dissipative drag force, the rheology is Newtonian \cite{MarschallStaples,MT1,OT-HB}. At low $\dot\gamma$ below $\phi_J$ one has $p\sim\dot\gamma$, while above $\phi_J$ one has \cite{OT-HB} a finite yield stress with $p\sim p_0+c\dot\gamma^b$.  Thus below $\phi_J$, $p/\dot\gamma$ should be independent of $\dot\gamma$ at sufficiently small $\dot\gamma$. 
For a given pair of strain rates $\dot\gamma_1<\dot\gamma_2$, we thus get a lower bound on $\phi_J$ from the largest packing $\phi$ at which the values of $p/\dot\gamma_1\approx p/\dot\gamma_2$.  We denote this lower bound as $\phi_1$.  As the values of $\dot\gamma_1$ and $\dot\gamma_2$ decrease, $\phi_1$ will increase towards $\phi_J$.
In Fig.~\ref{eta-vs-phi}a we plot $p/\dot\gamma$ vs $\phi$ for our smallest  strain rates $\dot\gamma=10^{-6}$ and $10^{-5}$, for crosses with aspect ratio $\beta=0.25$, 0.5 and 1.  The resulting lower bounds $\phi_1$ on $\phi_J$ are indicated by the dashed vertical lines in the figure.   

Another method often used to locate $\phi_J$ is to measure the average number of contacts per particle $Z$, and assert that jamming occurs when this reaches the isostatic value \cite{OHern}, for which the number of force constraints equals the number of degrees of freedom.  For frictionless particles this is $Z_\mathrm{iso}=2d_f$, where $d_f$ is the number of degrees of freedom per particle.  For 2D particles without rotational symmetry, $d_f=3$ and so $Z_\mathrm{iso}=6$.  However, for non-spherical particles it has been noted \cite{VanderWerf,MT1,Donev3,Donev4,Wouterse,Zeravcic,Mailman,Azema,Schreck,MTcompress} that jamming is often hypostatic, with $Z_J < Z_\mathrm{iso}$.  We thus expect that the value of $\phi$ at which $Z=Z_\mathrm{iso}$ gives an upper bound on $\phi_J$.  In Fig.~\ref{eta-vs-phi}b we plot $Z$ vs $\phi$ for our crosses; the dashed vertical lines indicate the values of $\phi$  where $Z=Z_\mathrm{iso}$.  We denote this upper bound by $\phi_{2}$.
We will take as our rough estimate for the jamming transition the average of this lower and upper bound, $\phi_J\approx(\phi_1+\phi_2)/2$.  In general we find the difference $\phi_2-\phi_1$ to be quite small.

{\color{black}Note, for computing $Z$ we count twice each contact where two spherocylinder segments touch side-to-side.  This is because each side-to-side contact constrains two degrees of freedom: the translational motion transverse to the contacting surface, as well as rotational motion \cite{VanderWerf,MTcompress,Azema3}.  This double counting of side-to-side contacts is a significant effect for elongated spherocylinders where there are many side-to-side contacts; however we find it to be a relatively small correction for the crosses.}

\begin{figure}
\resizebox{0.95\hsize}{!}{
\includegraphics{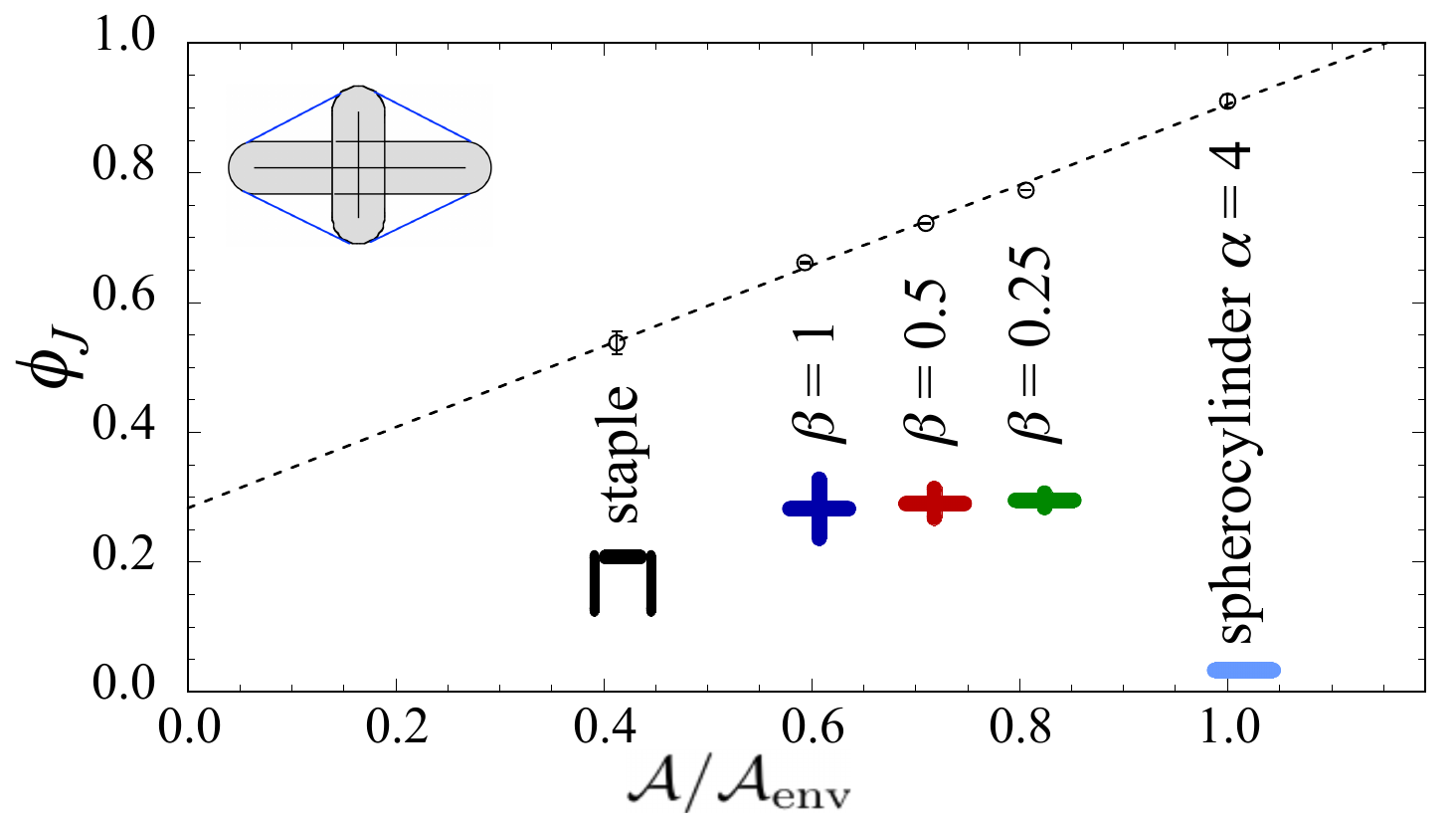}}
\caption{Estimated jamming transition $\phi_J$ for particles of different shape vs the ratio of particle area to the area of the particle's convex envelope $\mathcal{A}/\mathcal{A}_\mathrm{env}$.  
We show results for staples, crosses with aspect ratios $\beta=1$, 0.5 and 0.25, and spherocylinders with asphericity $\alpha=4$.
The upper limit of the error bar on each point represents the upper bound $\phi_2$ for $\phi_J$, determined by the condition $Z=Z_\mathrm{iso}=6$.  The lower limit of the error bars represents the lower bound $\phi_1$ for $\phi_J$, determined as the largest $\phi$ for which $p/\dot\gamma$ is the same for both $\dot\gamma=10^{-6}$ and $10^{-5}$.  
%For all points but the spherocylinder, the symbol is located at $\phi_J\approx (\phi_1+\phi_2)/2$; for the spherocylinder it is at a more precise value of $\phi_J$ determined by scaling methods [6].  
The dashed line is a linear fit to the data.  
{\color{black}In the inset shown in the upper left corner, the shaded gray region is the particle area $\mathcal{A}$, while the area bounded by the blue lines is the area of the convex envelope $\mathcal{A}_\mathrm{env}$.}
}
\label{phiJ-vs-Aenv}
\end{figure}

We now ask how $\phi_J$ varies with the particle shape.  To measure the degree of non-convexity of a particle we define the ratio,
$\mathcal{A}/\mathcal{A}_\mathrm{env}$,
where $\mathcal{A}$ is the area of the particle and $\mathcal{A}_\mathrm{env}$ is the area of the particle's convex envelop (see inset to Fig.~\ref{phiJ-vs-Aenv}).   In Fig.~\ref{phiJ-vs-Aenv} we plot our estimate for $\phi_J$ vs $\mathcal{A}/\mathcal{A}_\mathrm{env}$ for the crosses with $\beta=0.25$, $0.5$, and 1.  The upper limit of the error bars on the data points denotes the value of $\phi_2$, while the lower limit gives $\phi_1$.  For comparison we also include our earlier results for a system of $N=1024$ monodisperse staples \cite{MarschallStaples}  (as in Fig.~\ref{shapes}b), and for $N=1024$ spherocylinders of asphericity $\alpha=4$ \cite{MT1} (as in Fig.~\ref{shapes}a).  For the spherocylinders we use a bidisperse distribution of particle sizes to prevent spatial ordering, taking equal numbers of big and small particles with length scales in the ratio $D_b/D_s=1.4$, with $D_s=1$. 
{\color{black}We find a  linear dependence of $\phi_J$ on $\mathcal{A}/\mathcal{A}_\mathrm{env}$, with an extrapolated minimum value $\phi_J\approx 0.28$ as $\mathcal{A}/\mathcal{A}_\mathrm{env}\to 0$.  We may speculate that this minimum value might be achieved by crosses in the limit that the arms become infinitesimally thin.}

{\color{black}Note, this linear dependence should not be viewed as an exact relation; the data in Fig.~\ref{phiJ-vs-Aenv} are not exactly on a straight line, and for $\mathcal{A}/\mathcal{A}_\mathrm{env}=1$, which characterizes {\em all} convex particles, we know that there is a range of different possible values for $\phi_J$ depending on the particle shape (for example, $\phi_J=0.8433$ for bidisperse circles, as compared to $\phi_J\approx 0.906$ for bidisperse $\alpha=4$ spherocylinders \cite{MT1}).  We thus expect that, when including more diverse particle shapes, we will find a spread of values with a roughly linear in $\mathcal{A}/\mathcal{A}_\mathrm{env}$ trend, rather than an exact linear relation.}

{\color{black}Note also that the $\phi_J$ presented in Fig.~\ref{phiJ-vs-Aenv} are specifically for the case of shear-driven jamming, which is the subject of the present work.    The values of $\phi_J$ obtained for isotropic {\em compression-driven} jamming can be noticeably different.  The reason for this difference, for aspherical particles, can be attributed to the orientational ordering of particles that occurs under shear flow (see following section) but not under isotropic compression \cite{MarschallStaples,MT1,MT2}.

We can compare our results for $\phi_J$  with recent experiments on the isotropic and uniaxial compression of 2D $\beta=1$ crosses \cite{Zheng}.  In those experiments, the arms of the crosses are spherocylinders of $\alpha=5$ (compared to our particles with $\alpha=4)$;  these particles have  $\mathcal{A}/\mathcal{A}_\mathrm{env}= 0.519$.  If we use our results in Fig.~\ref{phiJ-vs-Aenv}, that would predict a $\phi_J\approx 0.61$.  The experiments, however, report the value  $\phi_J=0.475$.  There are several effects that might be responsible for the experimentally lower value of $\phi_J$:  (i) Our particles are frictionless while the experimental particles have inter-particle friction; such inter-particle friction generally lowers the jamming $\phi_J$ \cite{Makse,Otsuki}. (ii) As we speculated two paragraphs previously, the value of $\phi_J$ is likely not a simple function of $\mathcal{A}/\mathcal{A}_\mathrm{env}$ but may depend on other aspects of the particle's shape, perhaps the asphericity $\alpha$ of the arms of the cross. (iii) Our values of $\phi_J$ are for steady state simple shearing, while the experiment jams via isotropic and uniaxial compression. While uniaxial compression does contain a component of shear, isotropic compression does not shear the system and so likely does not induce any orientational ordering \cite{MarschallStaples,MT1,MT2}.  It may be that the different processes result in different values of $\phi_J$.  Despite the lack of agreement, it is noteworthy that both our simulations and the experiment find a $\phi_J$ that is significantly lower than the random close packing $\phi_J=0.8433$ for perfect 2D circles.
}

\begin{figure}
\resizebox{0.95\hsize}{!}{
\includegraphics{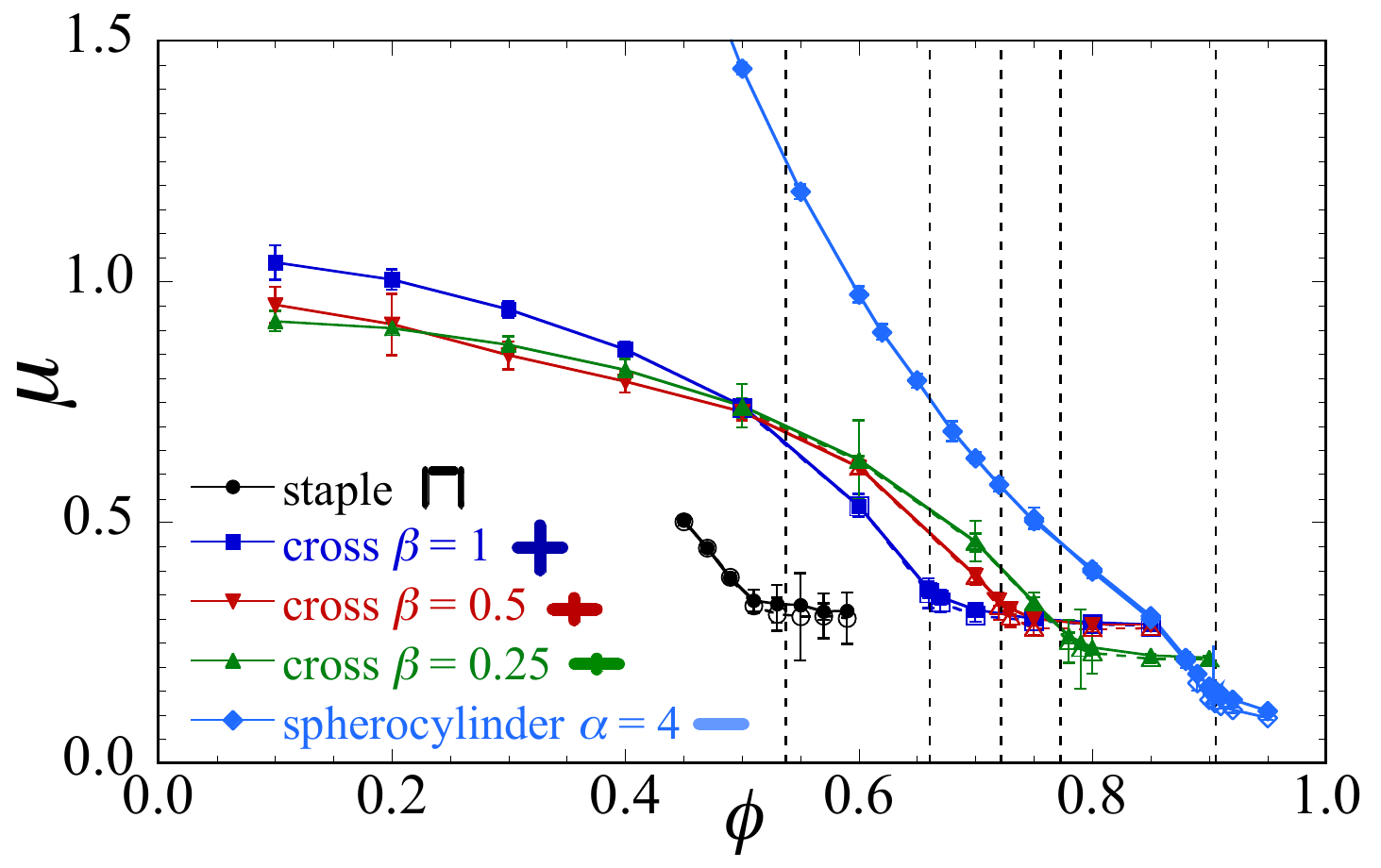}}
\caption{Macroscopic friction $\mu=\sigma_d/p$ vs packing $\phi$ for staples, crosses with aspect ratios $\beta = 1$, 0.5 and 0.25, and spherocylinders with asphericity $\alpha = 4$; vertical dashed lines locate the respective jamming transitions of these particles.  Solid symbols and solid lines are for $\dot\gamma=10^{-5}$ while open symbols and dashed lines are for $\dot\gamma=10^{-6}$, except for the staples where open symbols are for $\dot\gamma=5\times 10^{-6}$.
}
\label{mu-vs-phi}
\end{figure}

{\color{black}The jamming transition is also often characterized by the value of the macroscopic friction at the jamming point.  Although our particles have no microscopic inter-particle friction, the system as a whole does possess a macroscopic friction, defined as the ratio of the deviatoric shear stress $\sigma_d$ to the pressure $p$,
\begin{equation}
\mu=\sigma_d/p.
\end{equation}
Here $\sigma_d$ is defined in terms of the difference of eigenvalues of the ensemble averaged stress tensor of Eq.~(\ref{eP}),
\begin{equation}
\sigma_d=\sqrt{\langle p_{xy}\rangle^2 + \frac{1}{4}[\langle p_{xx}\rangle -\langle p_{yy}\rangle]^2}.
\end{equation}
In Fig.~\ref{mu-vs-phi} we plot this macroscopic friction $\mu$ vs packing $\phi$ for staples, our crosses of $\beta=1$, 0.5, and 0.25, as well as for spherocylinders of $\alpha=4$.  From this figure we determine the friction at jamming, $\mu_J$.
}

\begin{figure}
\resizebox{0.95\hsize}{!}{
\includegraphics{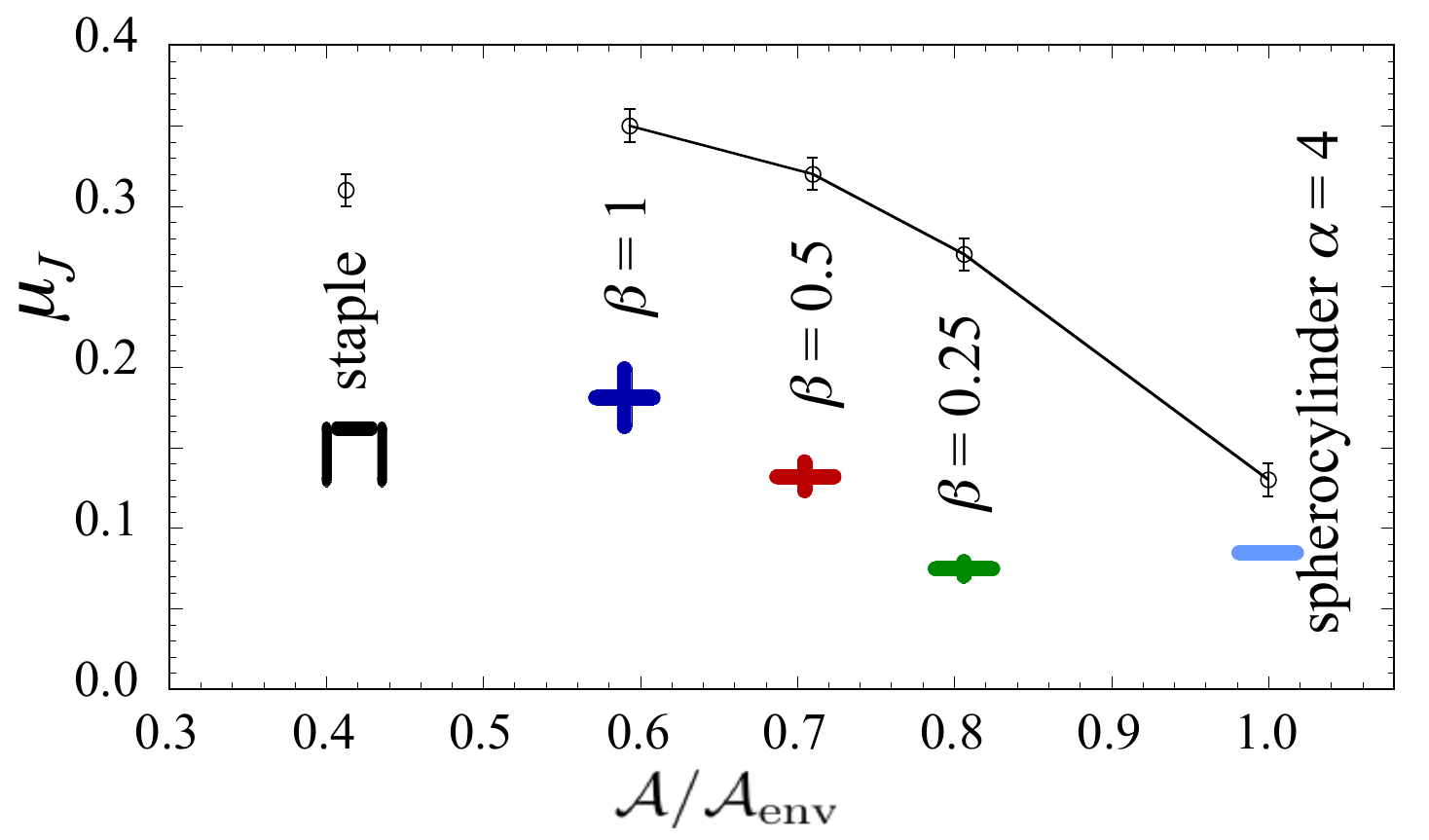}}
\caption{Macroscopic friction $\mu=\sigma_d/p$ vs $\mathcal{A}/\mathcal{A}_\mathrm{env}$ for staples, crosses with aspect ratios $\beta = 1$, 0.5 and 0.25, and spherocylinders with asphericity $\alpha = 4$.  $\mathcal{A}/\mathcal{A}_\mathrm{env}$ is the ratio of the particle's area to the area of the particle's convex envelope; the smaller is $\mathcal{A}/\mathcal{A}_\mathrm{env}$, the greater is the non-convexity of the particle's shape.
}
\label{muJ-vs-A}
\end{figure}

{\color{black}
In Fig.~\ref{muJ-vs-A} we plot $\mu_J$  vs the ratio $\mathcal{A}/\mathcal{A}_\mathrm{env}$ for the different shape particles.  Recall, $\mathcal{A}/\mathcal{A}_\mathrm{env}$ is the ratio of particle area to the area of the particle's convex envelop; as $\mathcal{A}/\mathcal{A}_\mathrm{env}$ decreases, the particle gets increasingly non-convex.
The spherocylinders, which can be viewed as  crosses with $\beta=0$, and the crosses with $\beta = 0.25$, 0.5 and 1, have values of $\mu_J$ that fall on a smooth curve; as $\mathcal{A}/\mathcal{A}_\mathrm{env}$ decreases, $\mu_J$ increases.  
We conjecture that, as the arms of the cross become more equal in length, the particles can more effectively interlock with each other, and so support a greater shear stress when jammed.  
For the staple we find that $\mu_J$ is still larger than for the convex spherocylinder, but smaller than the $\beta=1$ cross.  
For staples we have found \cite{MarschallStaples} that pairs of staples often nest one within the other to form an effectively convex square  composite particle; it may be that such composite particles more easily slide over one another than do the crosses, thus lowering $\mu_J$.}

\subsection{Rotational and Orientational Behavior}
\label{rot_orient}

\begin{figure}
\resizebox{0.95\hsize}{!}{
\includegraphics{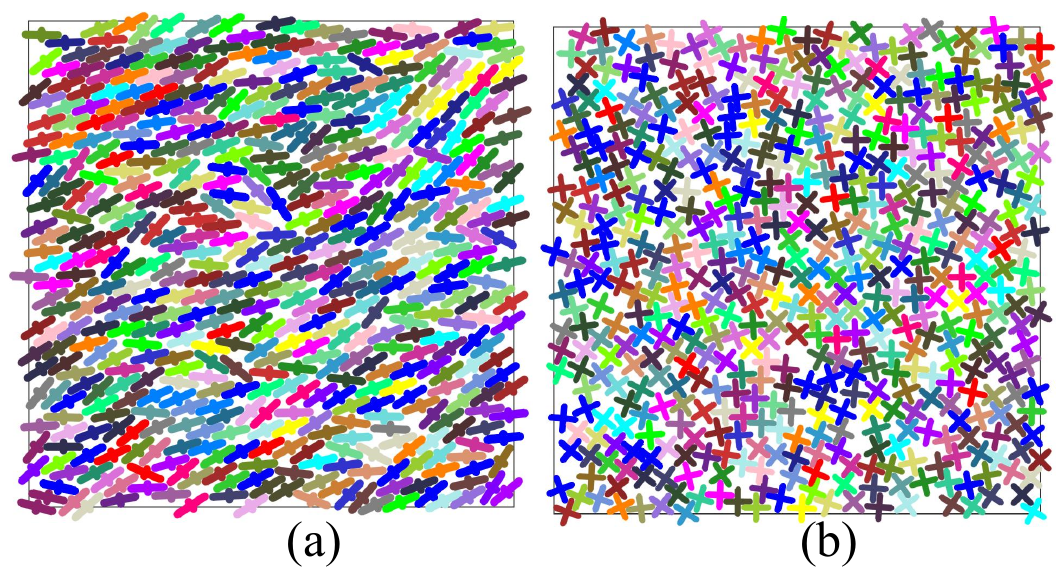}}
\caption{Sample configurations of $N=512$ crosses with (a) aspect ratio $\beta=0.25$ at packing $\phi=0.8$, and (b) $\beta=1$ at $\phi=0.7$; both are at strain rate $\dot\gamma=10^{-5}$.  Different colors are used to  distinguish different particles, but have no other meaning.  Animations of these two sheared configurations are shown as Online Resource 1 and Online Resource 2; we also show an animation of shearing for $\beta=0.5$ crosses at $\phi=0.75$ and $\dot\gamma=10^{-5}$ as Online Resource 3.
}
\label{crossConfigs}
\end{figure}

We now report our results for the average rotational motion and the orientational ordering of our particles.
In Fig.~\ref{crossConfigs} we show typical configurations sampled during steady-state shear.  In Fig.~\ref{crossConfigs}a we show crosses with $\beta=0.25$, as in Fig.~\ref{shapes}c, at a packing $\phi=0.8$ slightly above jamming ($\phi_J\approx 0.78$), at strain rate $\dot\gamma=10^{-5}$.  For this $\beta$ the short arm is such that only the semi-circular end caps protrude beyond the body of the long arm.  In Fig.~\ref{crossConfigs}b we show crosses with $\beta=1$, and so equal arm lengths as in Fig.~\ref{shapes}d, at a packing $\phi=0.7$ above jamming ($\phi_J\approx 0.67$), at $\dot\gamma=10^{-5}$.  By eye one sees the suggestion of nematic ordering in Fig.~\ref{crossConfigs}a with finite positive angle with respect to the flow direction $\mathbf{\hat x}$.  For the crosses with $\beta=1$ in Fig.~\ref{crossConfigs}b, the nematic ordering is necessarily zero due to the 4-fold rotational symmetry of the particle, and it is difficult to see whether there is any tetratic ordering or not.
We now quantify these observations.

\begin{figure}
\resizebox{0.95\hsize}{!}{
\includegraphics{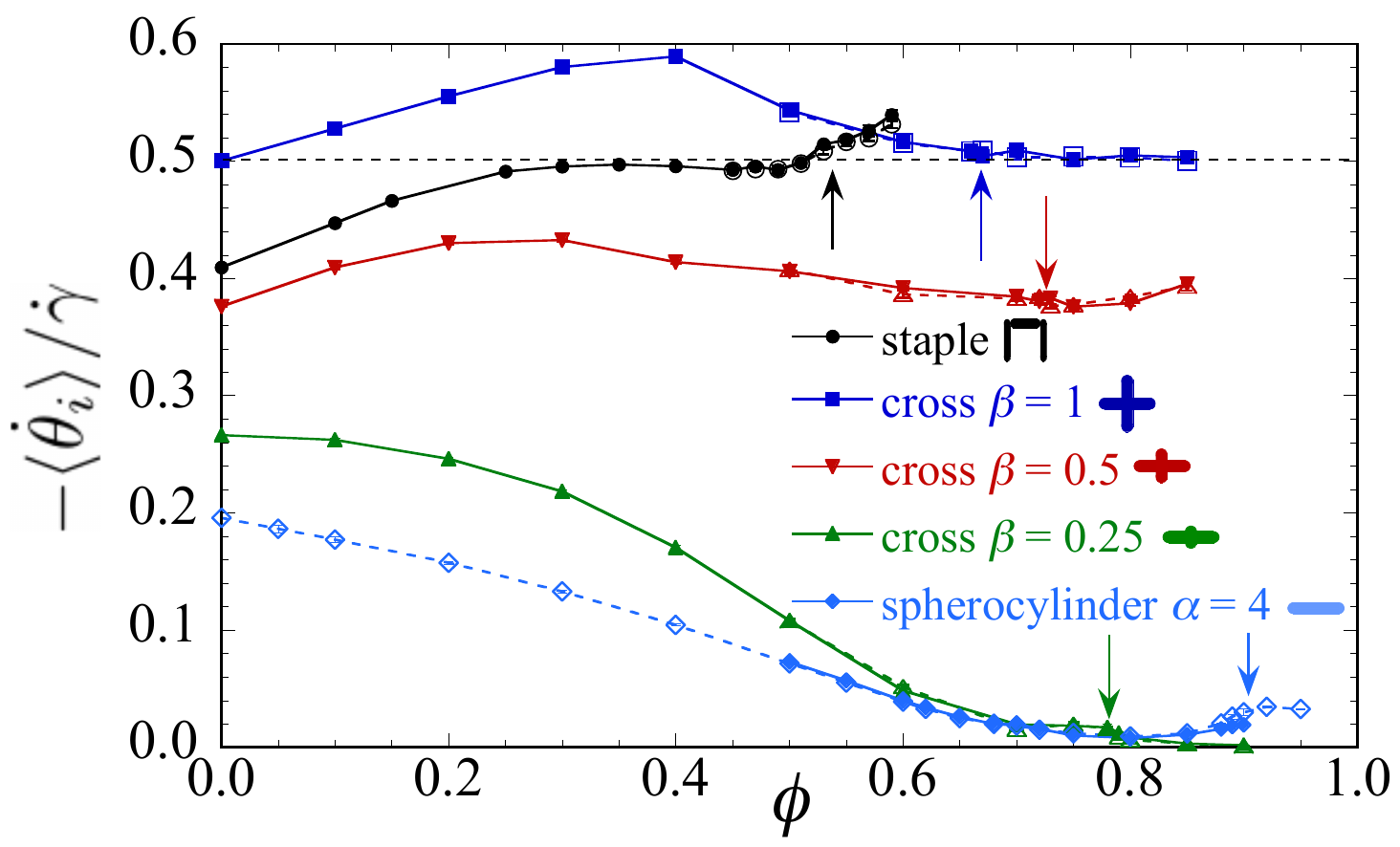}}
\caption{Scaled average angular velocity $-\langle\dot\theta_i\rangle/\dot\gamma$ vs packing $\phi$ for crosses with aspect ratio $\beta=0.25$, 0.5 and 1, compared to  spherocylinders of asphericity $\alpha=4$ and  staples.  For each case results are shown at a smaller strain rate $\dot\gamma_1$ (open symbols and dashed lines) and a larger strain rate $\dot\gamma_2$ (solid symbols and solid lines); see Table~\ref{tab1} for values.  The data points at $\phi=0$ are from Eq.~(\ref{eomega}) for isolated particles.  The horizontal dashed line indicates the value $-\langle\dot\theta_i\rangle/\dot\gamma=1/2$ for an isolated particle with 4-fold rotational symmetry.  Arrows give the approximate location of the jamming transition for each type of particle.
}
\label{omega-vs-phi}
\end{figure}

\begin{figure}
\resizebox{0.95\hsize}{!}{
\includegraphics{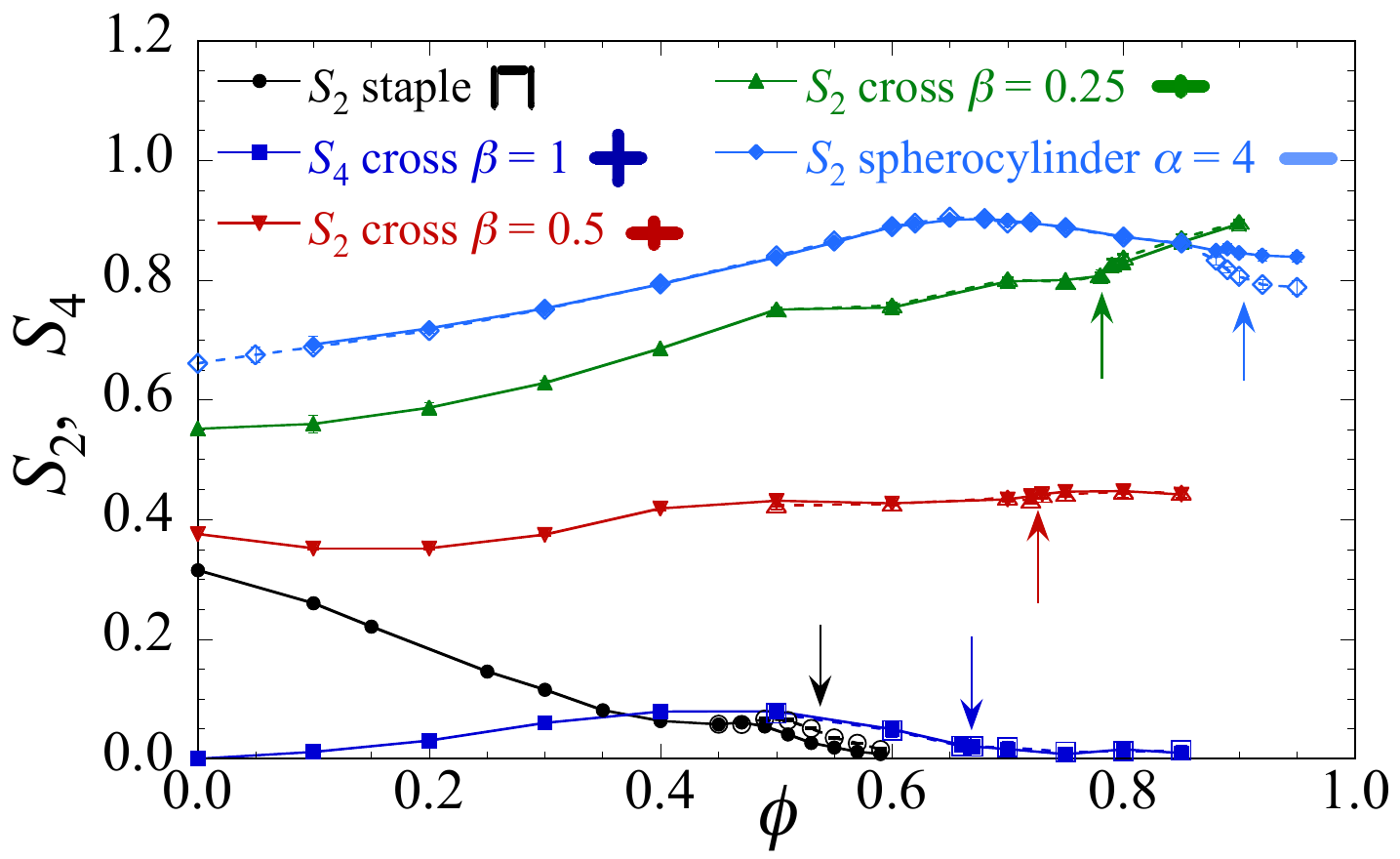}}
\caption{Magnitude of the orientational order parameter vs packing $\phi$ for crosses with aspect ratio $\beta=0.25$, 0.5 and 1, compared to  spherocylinders of asphericity $\alpha=4$ and  staples.  We show the nematic order parameter $S_2$ for all particles except the crosses with $\beta=1$, for which we show the tetratic order parameter $S_4$.
For each case results are shown at a smaller strain rate $\dot\gamma_1$ (open symbols and dashed lines) and a larger strain rate $\dot\gamma_2$ (solid symbols and solid lines).  The data points at $\phi=0$ are from Eq.~(\ref{eS2}) for isolated particles.  Arrows give the approximate location of the jamming transition for each type of particle.
}
\label{S2-vs-phi}
\end{figure}

In Fig.~\ref{omega-vs-phi} we plot the average particle rotational velocity, scaled by the strain rate, $-\langle\dot\theta_i\rangle/\dot\gamma$ vs $\phi$ for crosses with aspect ratio $\beta=0.25$, 0.5, and 1.   For comparison we include our earlier results for staples \cite{MarschallStaples} and spherocylinders with $\alpha=4$ \cite{MKOT,MT2}.
In each case, here and in subsequent Figs.~\ref{S2-vs-phi} and \ref{theta2-vs-phi}, we show results at both a smaller strain rate $\dot\gamma_1$ and a larger rate $\dot\gamma_2$ to illustrate that our strain rates are sufficiently small to be in the quasi-static limit, except possibly at the very densest packings.  The values of $\dot\gamma_1$ and $\dot\gamma_2$ are listed in Table~\ref{tab1}.
In Fig.~\ref{omega-vs-phi}, arrows denote the approximate location of the jamming transition $\phi_J$ for each particle shape.  It is  interesting that there is no clear signature of the location of $\phi_J$ from $-\langle\dot\theta_i\rangle/\dot\gamma$ and that in all cases $-\langle\dot\theta_i\rangle/\dot\gamma$ remains finite even in the dense configurations above jamming.

\begin{table}[h!]
\caption{Strain rate values used for data in Figs.~\ref{omega-vs-phi}, \ref{S2-vs-phi} and \ref{theta2-vs-phi}}
\begin{center}
\begin{tabular}{|l|c|c|}
\hline
shape & $\dot\gamma_1$ & $\dot\gamma_2$  \\
\hline
crosses& $10^{-6}$ & $10^{-5}$ \\
spherocylinders and staples & $10^{-5}$ & $10^{-4}$  \\
\hline
\end{tabular}
\end{center}
\label{tab1}
\end{table}%

For the convex spherocylinders we see that $-\langle\dot\theta_i\rangle/\dot\gamma$ rapidly decreases as $\phi$ increases, reaches a minimum, and then increases again as one approaches $\phi_J$.  We have observed similar behavior for spherocylinders of other asphericities $\alpha$ \cite{MKOT,MT2}.  The least non-convex of our non-convex particles, the cross with $\beta=0.25$, behaves qualitatively similar with an initial strong decrease, but then continues to monotonically decrease as $\phi$ goes above $\phi_J$.  
The other non-convex particles, however, have an initial increase as $\phi$ increases.  The staple reaches a plateau and then continues to increase, while the crosses with $\beta=0.5$ and 1 increase to a maximum and then decrease to values comparable to that of an isolated particle as $\phi$ goes above $\phi_J$.  
This increase can be thought of as a gear-like effect in which the interlocking of particles gives rise to torques that cause the particle to rotate faster than would an isolated particle.  In particular, the $\beta=1$ cross rotates faster than the value of 1/2 that represents the rotation of the affinely sheared host medium.

Next we consider the magnitude of the orientation ordering.  In Fig.~\ref{S2-vs-phi} we plot  the nematic order parameter $S_2$ vs $\phi$ for all particle shapes, except for the $\beta=1$ crosses where $S_2=0$ and so we show the tetratic $S_4$.  In general we see that $S_2$ is roughly anti-correlated with $-\langle\dot\theta_i\rangle/\dot\gamma$; when the latter is decreasing, $S_2$ is increasing, and vice versa.  Also, the shapes for which $-\langle\dot\theta_i\rangle/\dot\gamma$ is smallest tend to have the largest $S_2$.  Thus the slower the particles are rotating, the greater is the orientational ordering.  The only exception to this is the $\beta=1$ cross, where for the most part $S_4$ increases when $-\langle\dot\theta_i\rangle/\dot\gamma$ increases, and vice versa.  Orientational ordering is smallest for the $\beta=1$ cross, which in isolation (i.e., at $\phi=0$) shows no orientational ordering in the shear flow. 

\begin{figure}
\resizebox{0.95\hsize}{!}{
\includegraphics{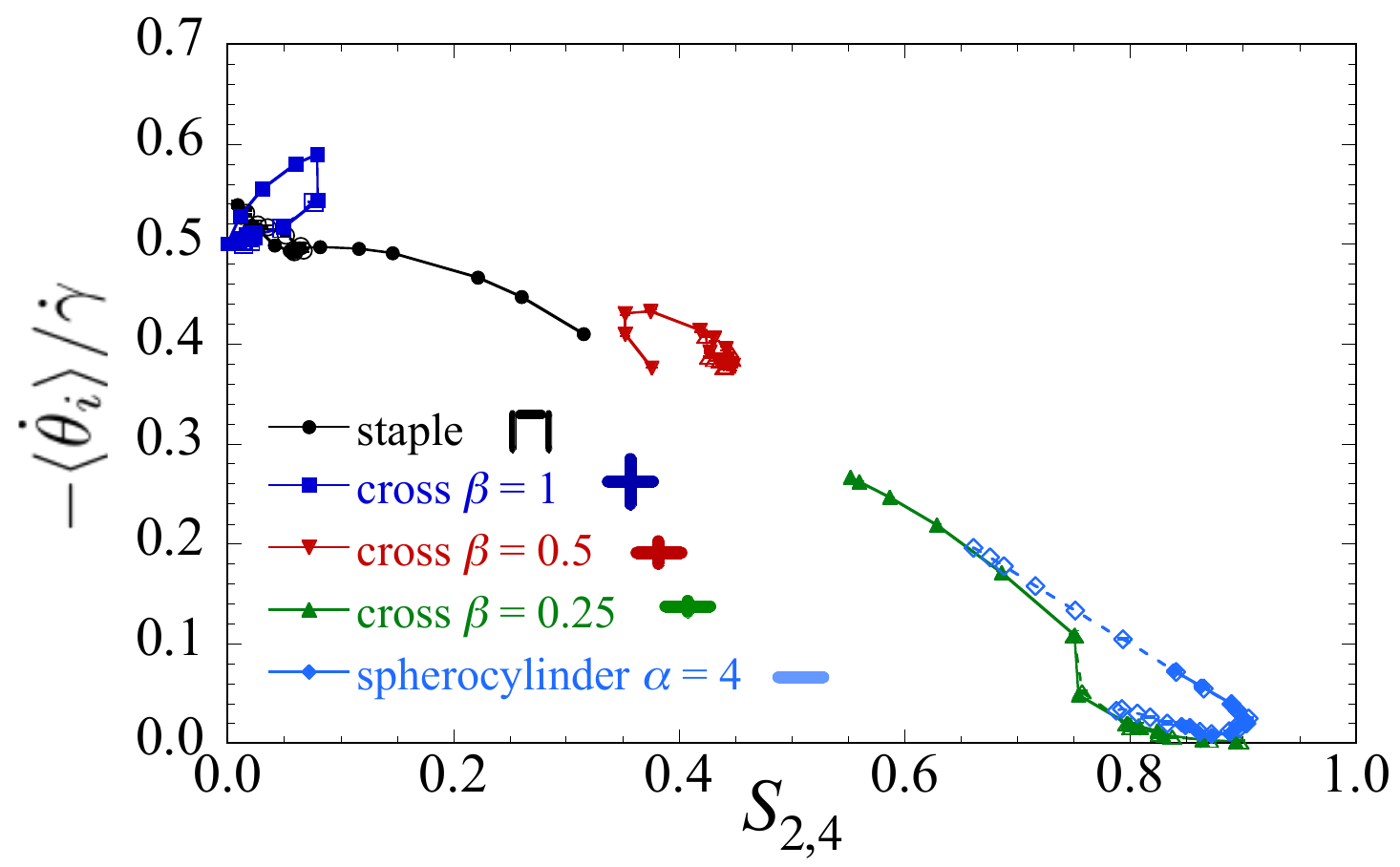}}
\caption{Parametric plot of $-\langle\dot\theta_i\rangle/\dot\gamma$ vs the magnitude of the nematic order parameter $S_2$, for 
crosses with aspect ratio $\beta=0.25$, and 0.5, compared to  spherocylinders of asphericity $\alpha=4$ and  staples.  
For  crosses with $\beta=1$, where $S_2=0$ by symmetry, we show a similar plot but using  the tetratic $S_4$.
The data is the same as in Figs.~\ref{omega-vs-phi} and \ref{S2-vs-phi}.  
For each case results are shown at a smaller strain rate $\dot\gamma_1$ (open symbols and dashed lines) and a larger strain rate $\dot\gamma_2$ (solid symbols and solid lines).  
}
\label{av-vs-S2}
\end{figure}

\begin{figure}
\resizebox{0.95\hsize}{!}{
\includegraphics{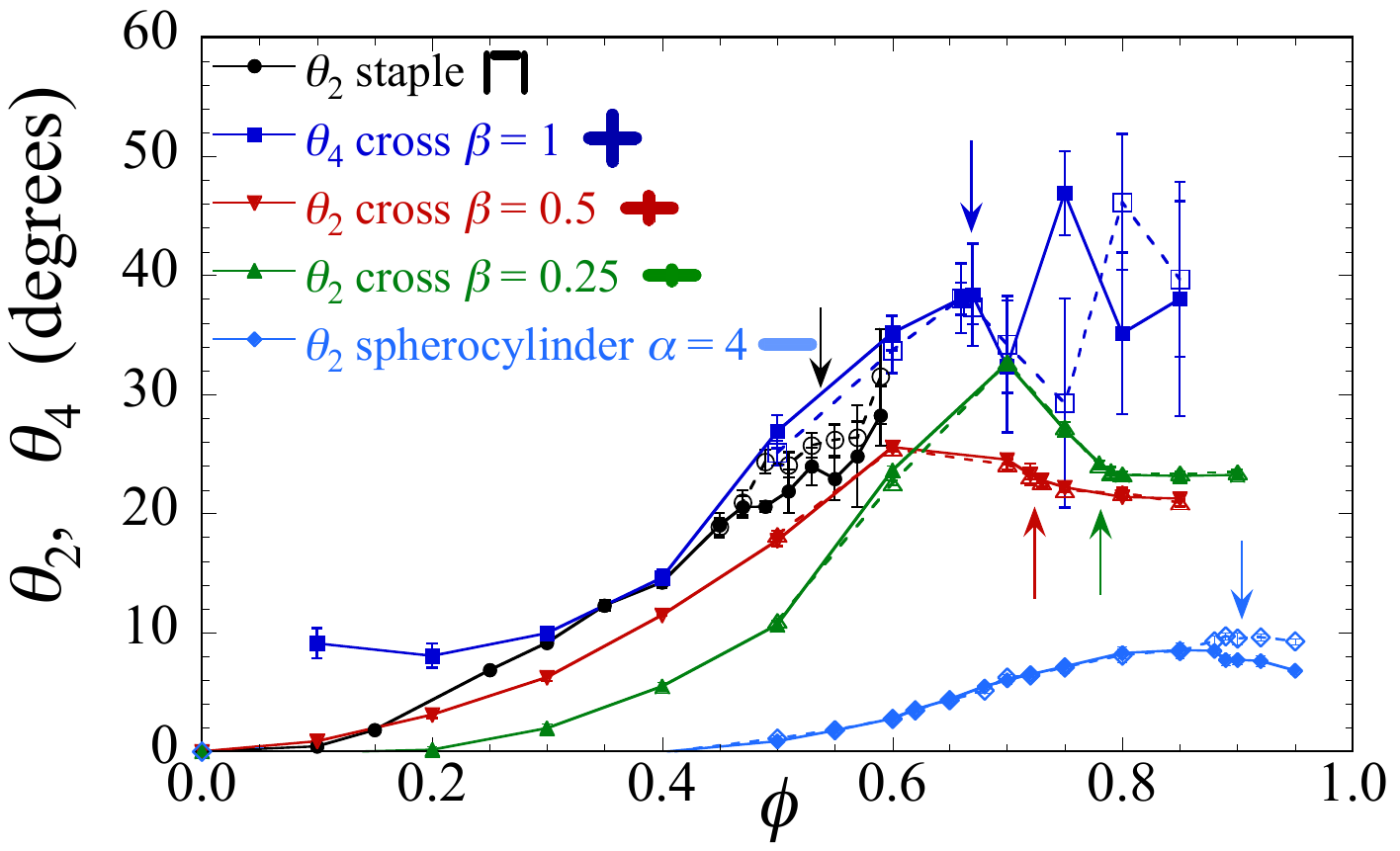}}
\caption{Direction of the orientational order parameter vs packing $\phi$ for crosses with aspect ratio $\beta=0.25$, 0.5 and 1, compared to  spherocylinders of asphericity $\alpha=4$ and  staples.  We show the nematic director orientation $\theta_2$ for all particles except the crosses with $\beta=1$, for which we show the tetratic orientation $\theta_4$.
For each case results are shown at a smaller strain rate $\dot\gamma_1$ (open symbols and dashed lines) and a larger strain rate $\dot\gamma_2$ (solid symbols and solid lines).  Arrows give the approximate location of the jamming transition for each type of particle.
}
\label{theta2-vs-phi}
\end{figure}

{\color{black}To highlight the relation between  $-\langle\dot\theta_i\rangle/\dot\gamma$ and $S_2$, in Fig.~\ref{av-vs-S2} we show a parametric plot of $-\langle\dot\theta_i\rangle/\dot\gamma$ vs $S_2$, using the same data as in Figs.~\ref{omega-vs-phi} and \ref{S2-vs-phi}.  For the $\beta=1$ crosses, where $S_2=0$ by symmetry, we use instead the tetratic $S_4$.  For all particles except the $\beta=1$ crosses, we see clearly the anti-correlation between
$-\langle\dot\theta_i\rangle/\dot\gamma$ and $S_2$;  $-\langle\dot\theta_i\rangle/\dot\gamma$ trends downwards as $S_2$ increases.  It is noteworthy that the curves for the different shapes seem to fall roughly around a common curve.  Only the $\beta=1$ crosses behave differently, showing a positive correlation between  $-\langle\dot\theta_i\rangle/\dot\gamma$ and $S_4$.}

Finally, in Fig.~\ref{theta2-vs-phi} we consider the direction of the orientational ordering, plotting the angle of the nematic director $\theta_2$ vs $\phi$ for all particle shapes, except for the $\beta=1$ cross where we show the tetratic $\theta_4$.  As has been well noted previously for spherocylinders and rod-shaped particles \cite{Guo1,Campbell,Guo2,Borzsonyi1,Borzsonyi2,Wegner,Wegner2,Trulsson,Nagy}, 
at finite density the particles orient at a finite positive angle $\theta_2>0$ with respect to the flow direction $\mathbf{\hat x}$, and this angle generally increases as the packing $\phi$ increases.  An interesting observation pertains to the $\beta=1$ cross.  All particles for which $\Delta I_i\ne 0$ orient with $\theta_2=0$ in the isolated particle limit, and this seems to be consistent with the $\phi\to 0$ behavior at finite packing shown in Fig.~\ref{theta2-vs-phi}.  However for the $\beta=1$ cross, which  has $\Delta I_i=0$, the isolated particle has a completely uniform orientation distribution $\mathcal{P}(\theta_i)=1/2\pi$, so $S_m=0$ for all $m$, and so $\theta_m$ in this isolated particle limit is undefined.
It is interesting, therefore, that we find in Fig.~\ref{theta2-vs-phi} that $\theta_4$ is finite and moreover appears to be approaching a finite, non-zero, value as $\phi\to 0$.

\subsection{Short Distance Correlations}
\label{correl}

\begin{figure}
\resizebox{0.95\hsize}{!}{
\includegraphics{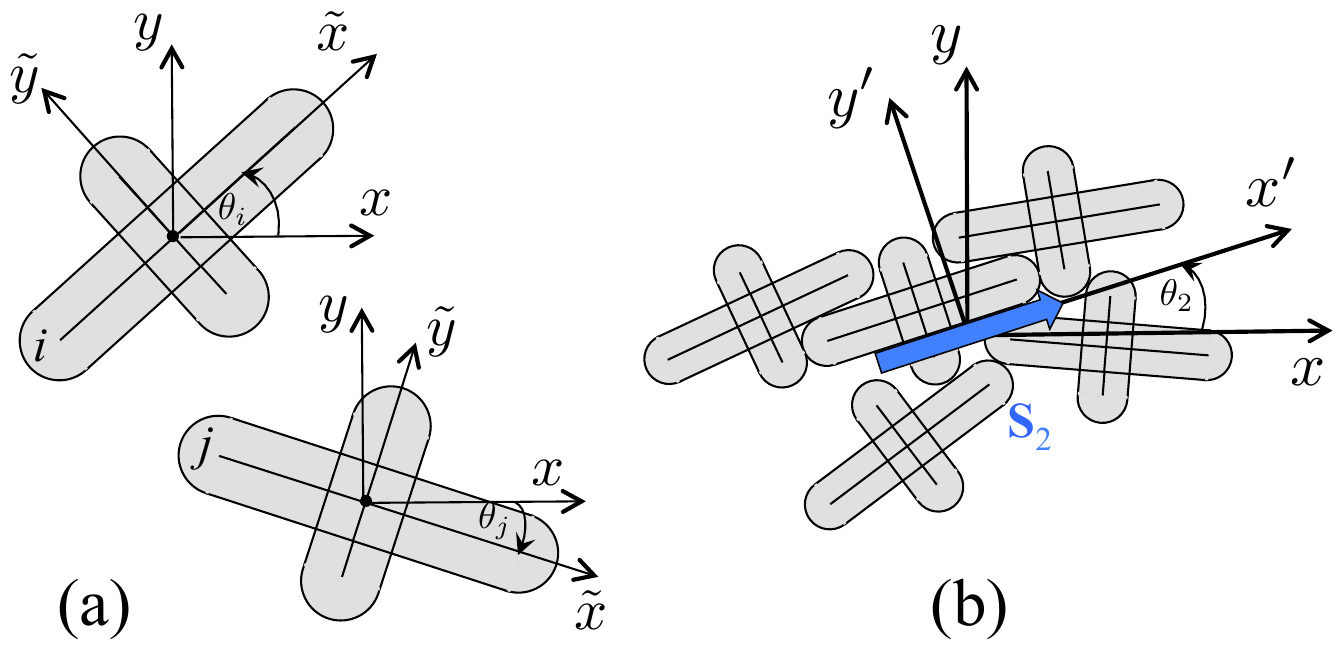}}
\caption{(a) Local coordinates $\tilde x$ and $\tilde y$ align with the long and short arms of a cross, and so vary from particle to particle. The axis $\tilde x$ on particle $i$ is at an angle $\theta_i$ with respect to the shear flow direction $\mathbf{\hat x}$.
(b) Global coordinates $x^\prime$ and $y^\prime$ align parallel and transverse to the direction of the global orientational order parameter $\mathbf{S}_2$, indicated by the large blue arrow.  The axis $x^\prime$ is at an angle $\theta_2$ with respect to the shear flow direction $\mathbf{\hat x}$.  Coordinates $x$ and $y$ are parallel and transverse to the direction of the shear flow.  
}
\label{cross-axes}
\end{figure}

To try to understand the above results concerning rotations and ordering, it is interesting to consider the correlations between neighboring particles in mutual contact, to see what is the geometry of local particle clusters.
Consider a given particle $i$, and construct a local coordinate system $(\tilde x,\tilde y)$ with origin at the center of mass $\mathbf{r}_i$, and the $\tilde x$ axis along the direction of the long arm, 
{\color{black}as illustrated in Fig.~\ref{cross-axes}a}.  
If $\mathbf{r}_j-\mathbf{r}_i$ is the displacement from the center of mass of particle $i$ to that of contacting particle $j$, we define the coordinates $(\tilde x, \tilde y) = (\mathbf{r}_j-\mathbf{r}_i)/[(D_i+D_j)/2]$ in this local coordinate system.  Note, for our monodisperse crosses, in which all particles have the same $D$, the denominator in this expression is just unity; but for our bidisperse spherocylinders the denominator rescales distances between different size particles to a common length.  We then define $\tilde g(\tilde x,\tilde y)$ as the probability density to find a contacting neighbor at $(\tilde x,\tilde y)$.  Similarly we define the orientational correlation $\tilde G_m(\tilde x,\tilde y)=\langle \cos(m[\theta_j-\theta_i])\rangle$, for particle $j$ at position $(\tilde x,\tilde y)$ with respect to $i$.  

Averaging over particles within a given configuration, and over configurations in our shearing ensemble, in Fig.~\ref{correl} we show intensity plots of $\tilde g(\tilde x,\tilde y)$ and $\tilde G_2(\tilde x,\tilde y)$ at packings $\phi\approx \phi_J$ for spherocylinders of asphericity $\alpha=4$, and crosses of aspect ratio $\beta=0.25$ and 0.5; for crosses of aspect ratio $\beta=1$ we show $\tilde g(\tilde x,\tilde y)$ and $\tilde G_4(\tilde x,\tilde y)$.   For $\tilde g(\tilde x,\tilde y)$ we use a logarithmic intensity scale to better highlight features.  For $\tilde G_2(\tilde x,\tilde y)$, dark blue denotes parallel oriented particles while dark red denotes perpendicular particles; white denotes particles with relative orientation of $\pi/4$.  For $\tilde G_4(\tilde x,\tilde y)$, dark blue denotes parallel or perpendicular particles, dark red denotes particles with relative orientation of $\pi/4$, and white denotes a relative orientation of $\pi/8$.
{\color{black}In both sets of correlations, we see an envelope surrounding the particle within which $\tilde g$ and $\tilde G_m$ vanish.  No other particle $j$ can have its center of mass position $\mathbf{r}_j$ within this envelope around particle $i$, without significant and unreasonable overlap between the particles.  We will refer to this as the ``excluded area," even though this definition is somewhat different from the standard definition of that term \cite{Onsager}.  Not surprisingly, as $\beta$ increases, this area increases and becomes a roughly more circular shape.}

For the spherocylinders, $\tilde g(\tilde x,\tilde y)$ and $\tilde G_2(\tilde x,\tilde y)$ show that the closest neighboring particles generally tend to be parallel, and that particles contacting along their mutual long flat sides at $\tilde y\approx 1$ tend have their point of contact smoothly distributed along the length of the flat side, with a slight peak in probability in the middle \cite{MT1}.  In contrast, for the crosses with $\beta=0.25$, the closest neighboring particles again tend to lie parallel, but the probability for particles making contact along the flat sides of their long arm at $\tilde y\approx 1$ have $\tilde g(\tilde x,\tilde y)=0$ for $-3<\tilde x<3$, since the long arm of one cross must butt up against the short arm of the other cross. 
Particles contacting at $\tilde y\approx 1.5$ have parallel long arms but the contacts tend to be between the long arm of one and the short arm of the other.   

\begin{figure}
\resizebox{0.95\hsize}{!}{
\includegraphics{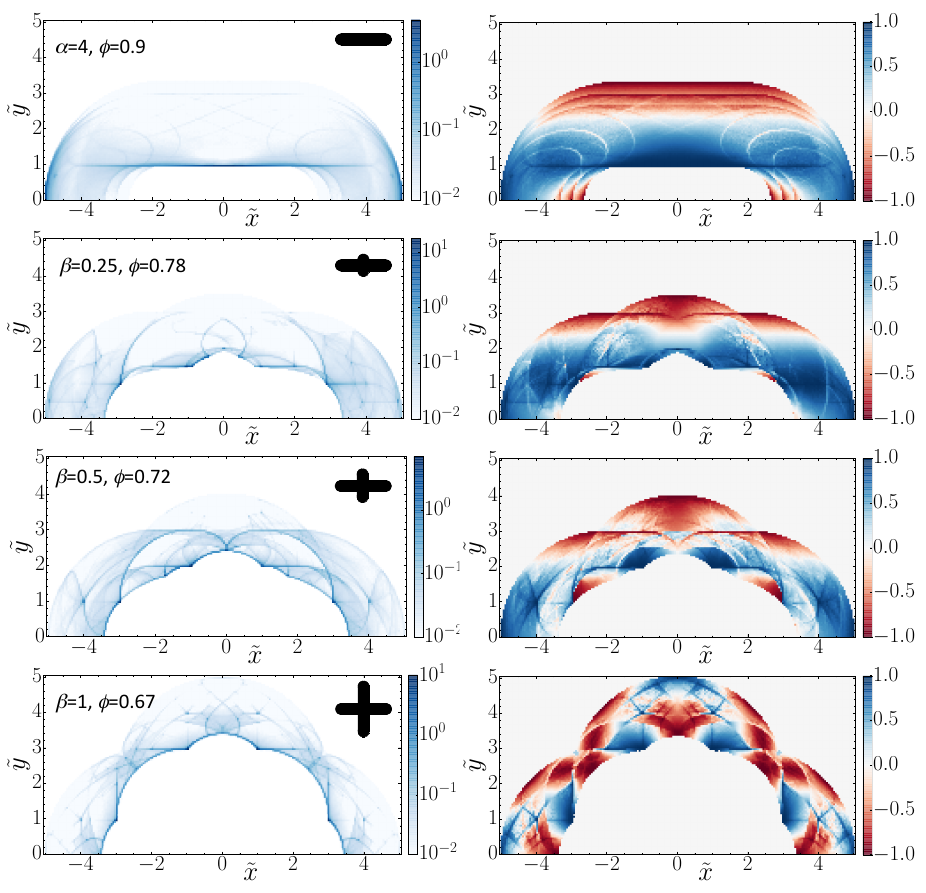}}
\caption{Intensity plots of spatial correlations $\tilde g(\tilde x,\tilde y)$ (left column, on logarithmic intensity scale) and orientational correlations $\tilde G_m(\tilde x,\tilde y)$ (right column) for spherocylinders of asphericity $\alpha=4$ (top row), and crosses with aspect ratio $\beta=0.25$, 0.5, and 1 (2nd, 3rd, and 4th rows) at packings $\phi\approx\phi_J$ near their respective jamming transitions.  We show the nematic ordering $\tilde G_2(\tilde x,\tilde y)$ for all but 
the $\beta=1$ cross, for which we show the tetratic $\tilde G_4(\tilde x,\tilde y)$.  The $(\tilde x,\tilde y)$ coordinate system is defined locally for each particle, with the $\tilde x$ axis taken along the direction of the long arm.  Icons in the upper right corner of the left column panels illustrate the particle shape of that row.
}
\label{correl}
\end{figure}

These observations may offer an explanation for our earlier result in Fig.~\ref{omega-vs-phi} that $-\langle\dot\theta_i\rangle/\dot\gamma$ in the dense region near, and going above, $\phi_J$  behaves differently for spherocylinders than for  crosses  $\beta=0.25$; while the former case shows an angular velocity that increases as $\phi$ increases, the latter case shows an angular velocity that monotonically decreases towards zero.
For the spherocylinders, the convex shape allows the particles to slide over each other as they shear, allowing greater freedom of motion.  For the $\beta=0.25$ crosses, $\tilde g(\tilde x,\tilde y)$ indicates a local structure of parallel but interlocking neighboring particles, with the short arms blocking such sliding motions, but being too small to exert sizable torques.

In contrast, the crosses with larger $\beta$ show a more complex pattern of neighbor orientations, oscillating between aligned and anti-aligned as one rotates around the particle.  As $\beta$, and hence $\mathcal{A}_\mathrm{env}/\mathcal{A}$, increases, it becomes harder to make dense packings in which particles are aligned, a conclusion consistent with the results of Fig.~\ref{S2-vs-phi}.

\subsection{Long Distance Correlations}
\label{long}

\begin{figure}
\resizebox{0.95\hsize}{!}{
\includegraphics{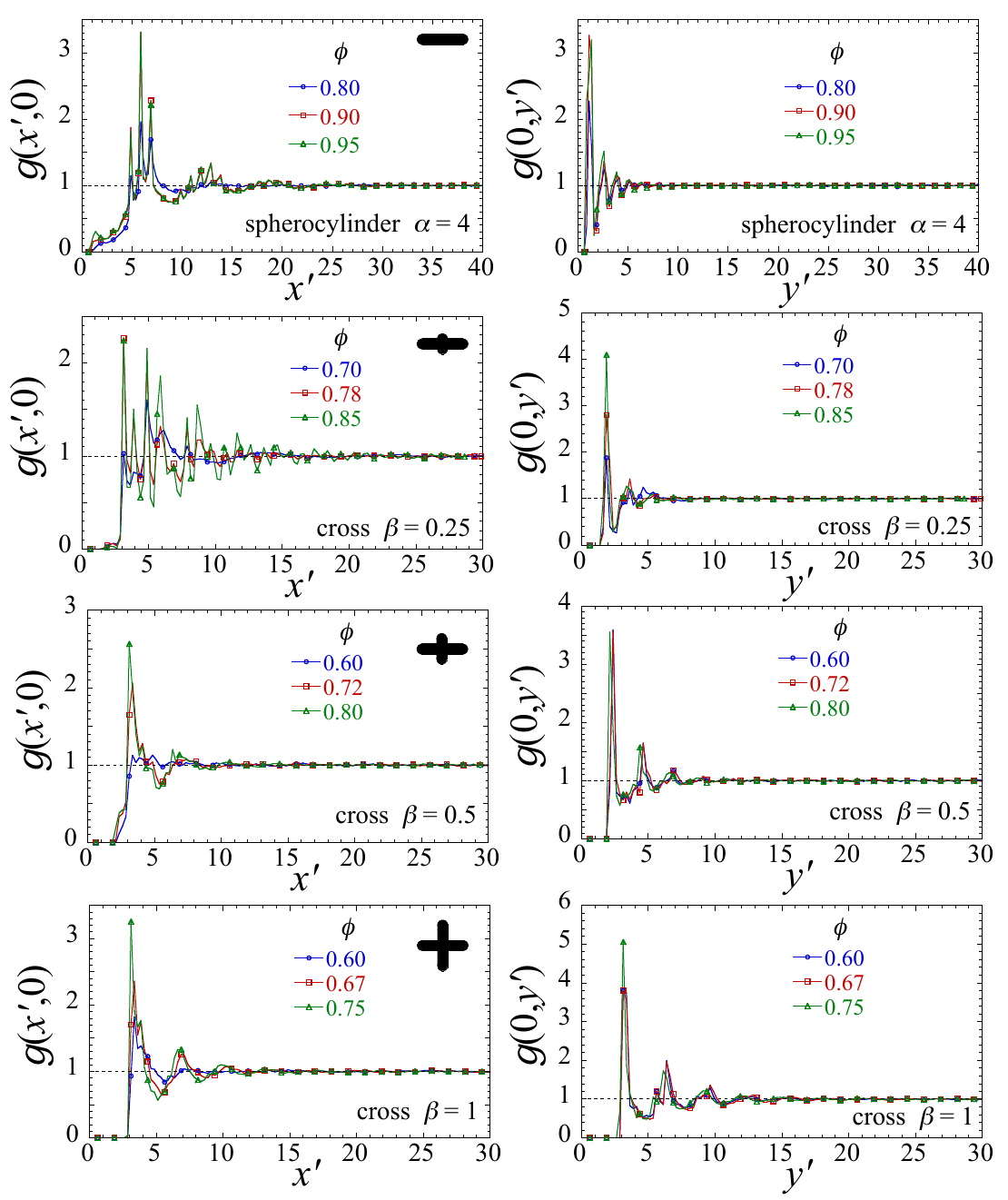}}
\caption{Pair correlation function $g(x^\prime,y^\prime)$ for spherocylinders with asphericity $\alpha=4$ (top row) and crosses of aspect ratio $\beta=0.25$, 0.5 and 1 (2nd, 3rd, and 4th rows).  Left hand column is $g(x^\prime, 0)$, right hand column is $g(0,y^\prime)$.
For each shape we show results at three values of the packing $\phi$: below $\phi_J$, roughly at $\phi_J$, and above $\phi_J$.
The axis $x^\prime$ lies in the direction of the nematic director (tetratic director for $\beta=1$).  Symbols are shown on every fifth data point.
Icons in the upper right corner of the left column panels illustrate the particle shape of that row.
}
\label{gxy}
\end{figure}

Finally we consider the long distance correlations in the system.  
%For circular particles there is no long range translational order in the system.  For non-circular particles, in which there is orientational ordering, it is interesting to see if anything is different.  
{\color{black}For circular particles in 2D, researchers generally  choose a bidisperse or polydisperse distribution of particle sizes to ensure that the system does not crystalize, and the absence of long range translational order has been confirmed in static jammed packings \cite{OHern}.  Sheared monodisperse frictionless spheres in 3D similarly show no translational order \cite{Sastry}.  For our size monodisperse  2D crosses, which show orientational ordering under shear, it is of interest to see if the orientational ordering induces any translational order.}
To investigate this we consider the pair correlation function, defined as usual,
\begin{equation}
g(\mathbf{r})= \frac{\mathcal{L}^2}{N}  \left\langle \frac{1}{N} \sum_{i\neq j} \delta(\mathbf{r} - \mathbf{r}_j+\mathbf{r}_i)\right\rangle.
\end{equation}
Because the system is sheared, correlations need not be isotropic.  So rather than showing a radial $g(r)$ averaged over separation directions, we instead consider $g(\mathbf{r})$ in two orthogonal directions.  We define the $x^\prime$ axis to be parallel to the nematic order parameter $\mathbf{S}_2$, at angle $\theta_2$ with respect to the flow direction (or at angle $\theta_4$ along the tetratic order parameter $\mathbf{S}_4$ for $\beta=1$ crosses), and $y^\prime$ as the orthogonal direction, 
{\color{black}as illustrated in Fig.~\ref{cross-axes}b}.  
In Fig.~\ref{gxy} we then plot $g(x^\prime,0)$ (left column) and $g(0,y^\prime)$ (right column) for spherocylinders with $\alpha=4$ (top row) and crosses of $\beta=0.25$, 0.5 and 1 (2nd, 3rd and 4th rows).  For each case we show results at three different packings $\phi$, one below $\phi_J$, one roughly at $\phi_J$ and one above $\phi_J$.  
In all cases we see a rapid decay to the large distance limit $g(\infty)=1$.  
{\color{black}For the range of $\phi$ shown, the systems of spherocylinders and crosses of $\beta=0.25$, 0.5 and 1 have lengths $\mathcal{L}\approx 90$, 60, 68, and 80 respectively, {\color{black}where lengths are measured in units of the spherocylinder width $D$} (for spherocylinders we use a system of twice the size as the crosses, hence the larger $\mathcal{L}$; for the crosses, $\mathcal{L}$ increases as 
the $\phi$ in the figures decreases).  We thus see that the decay to  $g\to 1$ occurs well before we reach the length scale of the system size.}
Thus, as with circular particles, there is no long range translational order, and the length scale of the decay does not seem to vary appreciably with $\phi$.

{\color{black}For a system of hard spherical particles of diameter $D$, the nearest any two particles may approach each other is $D$, and so the pair correlation $g(\mathbf{r})$ has a sharp jump from zero at $|\mathbf{r}|=D$; as $\phi\to\phi_J$ from below, the height of this peak at $|\mathbf{r}|=D$ diverges \cite{OHern}.  For aspherical particles, this nearest possible distance depends on the relative orientation of the two particles, and so behavior can be more complex.  This is most readily seen for the spherocylinders of $\alpha=4$, shown in the top row of Fig.~\ref{gxy}.  Consider the local particle based coordinates $(\tilde x,\tilde y)$, defined in Fig.~\ref{cross-axes}a.  From Fig.~\ref{correl} we see that the spherocylinder can have no other particle closer than $\tilde x=3$, if one looks for contacts in the direction parallel to the spherocylinder spine.  But if one looks in the transverse direction, one finds that another particle can be as close as $\tilde y=1$; this corresponds to two aligned spherocylinders, one lying on top of the other.  Since the global $(x^\prime, y^\prime)$ coordinates are parallel and transverse to the direction of the nematic order parameter $\mathbf{S}_2$, and since the spherocylinders are on average aligned with $\mathbf{S}_2$, it is therefore not surprising that $g(0,y^\prime)$ for the spherocylinder takes a sharp increase from zero at $y^\prime=1$.  One might then expect that $g(x^\prime, 0)$ should vanish for $x^\prime < 3$, however we see this is not so.  This is because all particles are not  aligned  exactly parallel to $\mathbf{S}_2$; if a particle happened to be aligned perpendicular to $\mathbf{S}_2$ (so that the local coordinate $\tilde y$ is aligned with the global coordinate $x^\prime$), it could then be in contact with another particle that is only a distance $\tilde y=x^\prime=1$ away.  The probability for this perpendicular alignment is small, which is why $g(x^\prime,0)$ takes a rather gradual increase above zero as $x^\prime$ increases above unity, unlike the sharp jump seen for $g(0,y^\prime)$ at $y^\prime=1$.  In contrast, for the crosses,  we see in Fig.~\ref{correl} that as $\beta$ increases, the excluded area becomes more circular and so the difference between $\tilde x$ and $\tilde y$ becomes less signifiant, and hence the difference between $x^\prime$ and $y^\prime$ becomes less noticeable, and so for $\beta=1$ we see that $g(x^\prime,0)$ and $g(0,y^\prime)$ both vanish for $x^\prime, y^\prime \lesssim 3$.
}

\begin{figure}
\resizebox{0.95\hsize}{!}{
\includegraphics{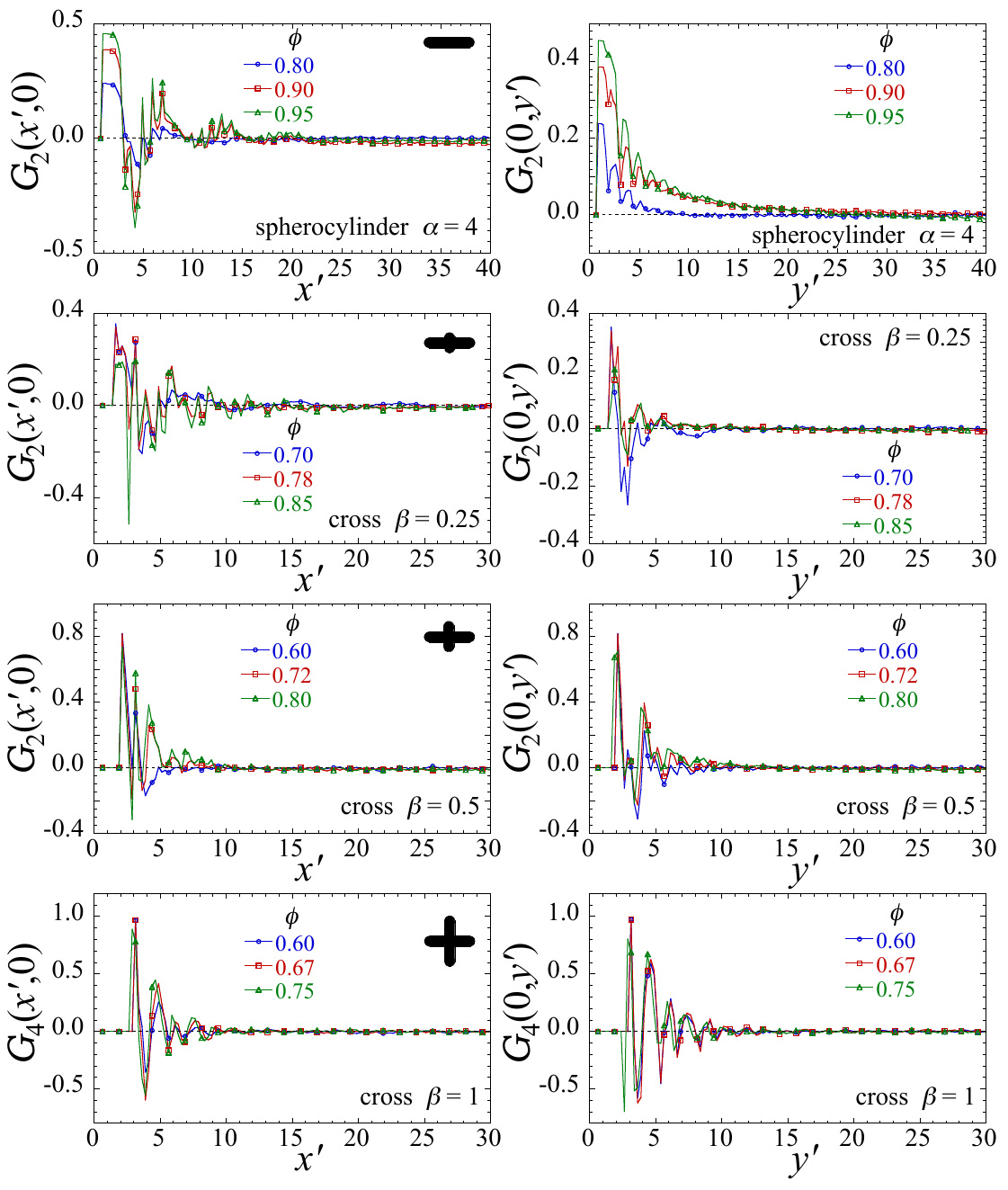}}
\caption{Nematic correlation function $G_2(x^\prime,y^\prime)$ for spherocylinders with asphericity $\alpha=4$ (top row) and crosses of aspect ratio $\beta=0.25$ and 0.5 (2nd and 3rd rows); tetratic correlation $G_4(x^\prime,y^\prime)$ for $\beta=1$ (4th row).  Left hand column is $G_{2,4}(x^\prime, 0)$, right hand column is $G_{2,4}(0,y^\prime)$.
For each shape we show results at three values of the packing $\phi$: below $\phi_J$, roughly at $\phi_J$, and above $\phi_J$.
The axis $x^\prime$ lies in the direction of the nematic director (tetratic director for $\beta=1$).  Symbols are shown on every fifth data point.
Icons in the upper right corner of the left column panels illustrate the particle shape of that row.
}
\label{G2xy}
\end{figure}

Next we consider the correlations of the orientational order.  Defining this correlation function as,
\begin{equation}
G_m(\mathbf{r})=\left\langle \cos(m[\theta_j-\theta_i])\right\rangle -S_m^2,
\end{equation}
for particle $j$ at position $\mathbf{r}=\mathbf{r}_j-\mathbf{r}_i$ with respect to particle $i$,
in Fig.~\ref{G2xy} we  plot the nematic correlation $G_2(x^\prime,0)$ (left column) and $G_2(0,y^\prime)$ (right column) for spherocylinders with $\alpha=4$ (top row) and crosses of $\beta=0.25$ and 0.5 (2nd, 3rd rows); for crosses with $\beta=1$ we plot the tetratic $G_4$ (4th row).  For each case we show results at three different packings $\phi$, one below $\phi_J$, one roughly at $\phi_J$ and one above $\phi_J$.  In all cases we see a rapid decay to the large distance limit $G_m(\infty)=0$, and little dependence of the decay length on $\phi$.  The only exception is for the transverse correlation $G_2(0,y^\prime)$ for the spherocylinders, where we see a noticeable increase in the decay length as $\phi$ increases to jamming, but this length nevertheless remains finite.  We thus conclude that there are no long range orientational correlations in the system, and we infer that it is the shearing that acts like a finite ordering field for the orientational order, rather than the orientational order being the result of a many particle collective behavior.  We found similar results previously for staples \cite{MarschallStaples}.

\section{Conclusions}
\label{conclusions}

We have considered the shear-driven flow of athermal, frictionless, non-convex cross-shaped particles of varying aspect ratios in uniform steady-state, focusing on the rotational motion of the particles and orientational ordering.  We have compared our results with our prior results from convex spherocylinders and non-convex staple-shaped particles.  Comparing particles of different shape, we find  the novel result that the jamming transition $\phi_J$ seems  to scale  linearly with the ratio of the particle's area to the area of the particle's convex envelope, $\mathcal{A}/\mathcal{A}_\mathrm{env}$. 

Considering rotational motion, we find that the particle angular velocity and the orientational ordering depend sensitively on the shape of the particle.  For convex spherocylinders we find that the scaled average angular velocity $-\langle \dot\theta_i\rangle/\dot\gamma$ always initially decreases as the particle density is increased from the isolated particle limit; collisions tend to slow rotation.  However upon further increasing the packing $\phi$, $-\langle \dot\theta_i\rangle/\dot\gamma$ reaches a finite minimum and then increases as the jamming transition is approached.  This is true for spherocylinders of any asphericity $\alpha$ \cite{MKOT,MT2}.  In contrast, for strongly non-convex crosses we find that $-\langle \dot\theta_i\rangle/\dot\gamma$ increases as the  particle density is increased from the isolated particle limit; collisions tend to increase rotation.  Upon further increasing $\phi$, $-\langle \dot\theta_i\rangle/\dot\gamma$ reaches a finite maximum and then decreases as the jamming transition is approached.  For staples we find that $-\langle \dot\theta_i\rangle/\dot\gamma$ increases, then plateaus, then increases again as $\phi$ increases.

The magnitude of the nematic order parameter $S_2$ is in general similarly non-monotonic in the packing $\phi$, and appears to be anti-correlated with the angular velocity.  When $-\langle \dot\theta_i\rangle/\dot\gamma$ is large, $S_2$ is small, and vice versa.  An interesting exception is the case of the $\beta=1$ cross which has 4-fold rotational symmetry and so, when in isolation, rotates with a uniform $-\langle \dot\theta_i\rangle/\dot\gamma=1/2$, just as would a circular particle; the isolated particle shows no orientational ordering.  In this case we find that the interaction between particles at finite density leads to a small but finite tetratic ordering $S_4$, and  that $-\langle \dot\theta_i\rangle/\dot\gamma$ and the tetratic order $S_4$ are positively correlated; as $\phi$ increases, both quantities increase, reach a maximum, then decreases.  In the intermediate $\phi$ region, $-\langle \dot\theta_i\rangle/\dot\gamma>1/2$ is larger than the angular velocity of the affinely sheared host medium.

{\color{black}It is interesting to compare the behavior of  spherocylinders with that of crosses with aspect ratio $\beta=0.25$.  For such crosses, the short arm appears as two semicircular bumps on the opposite sides of the otherwise flat sides of the long arm, as shown in Fig.~\ref{shapes}(c). One can therefore view such a cross as a spherocylinder with an asperity on each flat side that inhibits sliding motion along these sides, thus making an analogy between a $\beta=0.25$ cross and a spherocylinder with inter-particle frictional interactions.  Indeed, the decoration of convex particle surfaces with such asperities \cite{Papanikolaou}, or the rigid attachment of convex particles into a non-convex shape \cite{Buchholtz,Alonso,Torres}, have been previously used as models for frictional particles.
In our case, since the asperity formed by the short arm can withstand a large transverse force, and so provide an effective large tangential force against sliding along the long arm, we should view our $\beta=0.25$ cross as a spherocylinder with a very large coefficient $\mu_p$ of inter-particle friction.

To contrast the behavior of our frictionless spherocylinders with the $\beta=0.25$ crosses, to see how well the $\beta=0.25$ cross may indeed be behaving like a frictional spherocylinder, we compare with recent simulations of sheared 2D frictional ellipses \cite{Trulsson} in which a standard Cundall-Strack \cite{CS} form for the tangential Coulombic friction is used.  Considering Fig.~6(a) of Ref.~\cite{Trulsson} one sees that for a fixed ellipse aspect ratio, the jamming packing fraction $\phi_J$ decreases as the inter-particle friction coefficient $\mu_p$ increases, a result well known for spherical particles \cite{Makse,Otsuki}.  Taking the case of their most elongated ellipses and comparing the frictionless $\mu_p=0$ case against their most frictional  $\mu_p=10$ case, one finds a reduction in $\phi_J$ by roughly a factor of 0.87.  This compares reasonably well with the reduction in $\phi_J$ by a factor of 0.85 that we see in Fig.~\ref{phiJ-vs-Aenv}, comparing frictionless spherocylinders with $\beta=0.25$ crosses.  

From Fig.~6(e) of Ref.~\cite{Trulsson} we see that the macroscopic friction at jamming $\mu_J$ increases as the inter-particle friction coefficient $\mu_p$ increases, reaches a maximum near $\mu_p\approx 1$, then decreases to a limiting value that is still well above the frictionless case.  Comparing their frictionless $\mu_p=0$ case against their most frictional $\mu_p=10$ case, for their most elongated ellipses, gives an increase of $\mu_J$ by a factor of roughly 7. This compares with the increase in $\mu_J$ by a factor of roughly 2 that we see in Fig.~\ref{muJ-vs-A}, comparing our frictionless spherocylinders with $\beta=0.25$ crosses.  Thus, while the magnitudes are not in such good agreement, still the trend is the same; the presence of an asperity on the flat side of a spherocylinder increases the macroscopic friction $\mu_J$.

Finally, another feature observed when shearing strongly frictional particles is that the jamming transition appears to be discontinuous, with a finite jump in the yield stress at jamming \cite{Otsuki}.  However our results in Fig.~\ref{eta-vs-phi} give no suggestion of any such discontinuity.  Thus we conclude that, while the $\beta=0.25$ cross shares some characteristics of a frictional spherocylinder, in other respects this analogy remains lacking.  We had arrived at a similar conclusion in our earlier work on the jamming of frictionless staples \cite{MarschallStaples}.
}

The results discussed in this work clearly illustrate that particle shape, and in particular the degree of non-convexity, can lead to qualitative differences in the 
rotational motion and orientational ordering of non-spherical particles in a uniform shear driven flow, {\color{black}and to properties at the jamming transition}.

\section*{Compliance with ethical standards}

The authors declare that they have no conflict of interests.

\section*{Acknowledgements}
\label{ack}

This article is dedicated to the memory of Robert Behringer, who was much interested in the jamming of crosses, stars, and other oddly shaped granular particles.
{\color{black}We thank an anonymous reviewer for several helpful and interesting comments.}
This work was supported in part by National Science Foundation Grants CBET-1435861 
{\color{black}and DMR-1809318}. Computations were carried out at the Center for Integrated Research Computing at the University of Rochester.

\end{document}